\documentclass[12pt]{article}
\usepackage{amsfonts}
\font\dynkfont=cmsy10 scaled\magstep4    \skewchar\dynkfont='60
\def\dynk{\textfont2=\dynkfont}
\def\hr#1,#2;{\dimen0=.4pt\advance\dimen0by-#2pt
              \vrule width#1pt height#2pt depth\dimen0}
\def\vr#1,#2;{\vrule height#1pt depth#2pt}
\def\blb#1#2#3#4#5
            {\hbox{\ifnum#2=0\hskip11.5pt
                   \else\ifnum#2=1\hr13,5.4;\hskip-1.5pt
                   \else\ifnum#2=2\hr13.5,7.6;\hskip-13.5pt
                                  \hr13.5,3.2;\hskip-2pt
                   \else\ifnum#2=3\hr13.7,8.4;\hskip-13.7pt
                                  \hr13,5.4;\hskip-13pt
                                  \hr13.7,2.4;\hskip-2.2pt
                   \else\ifnum#2=7\hr13.7,8.4;\hskip-13.7pt
                                  \hr13.2,6.4;\hskip-13.2pt
                                  \hr13.2,4.4;\hskip-13.2pt
                                  \hr13.7,2.4;\hskip-2.2pt
                   \else\ifnum#2=8\hr13.7,8.4;\hskip-13.7pt
                                  \hr13.2,6.4;\hskip-13.2pt
                                  \hr13.2,4.4;\hskip-13.2pt
                                  \hr13.7,2.4;\hskip-8.2pt$\rangle$\hskip-2.0pt
                   \else\ifnum#2=4\hr13.5,7.6;\hskip-13.5pt
                                  \hr13.5,3.2;\hskip-8pt$\rangle$\hskip-1.8pt
                   \else\ifnum#2=5\hr30,5.4;\hskip-1.5pt         
                   \else\ifnum#2=6\hr13.7,8.4;\hskip-13.7pt
                                  \hr13,5.4;\hskip-13pt
                                  \hr13.7,2.4;\hskip-8.2pt$\rangle$\hskip-2.0pt
                      \fi\fi\fi\fi\fi\fi\fi\fi\fi
                   $#1$
                   \ifnum#4=0
                   \else\ifnum#4=1\hskip-9.2pt\vr22,-9;\hskip8.8pt
                   \else\ifnum#4=2\hskip-10.9pt\vr22,-8.75;\hskip3pt
                                  \vr22,-8.75;\hskip7.1pt
                   \else\ifnum#4=3\hskip-12.6pt\vr22,-8.5;\hskip3pt
                                  \vr22,-9;\hskip3pt
                                  \vr22,-8.5;\hskip5.4pt
                   \else\ifnum#4=5\hskip-9.2pt\vr39,-9;\hskip8.8pt
                                  \fi\fi\fi\fi\fi
                   \ifnum#5=0
                   \else\ifnum#5=1\hskip-9.2pt\vr1,12;\hskip8.8pt
                   \else\ifnum#5=2\hskip-10.9pt\vr1.25,12;\hskip3pt
                                  \vr1.25,12;\hskip7.1pt
                   \else\ifnum#5=3\hskip-12.6pt\vr1.5,12;\hskip3pt
                                  \vr1,12;\hskip3pt
                                  \vr1.5,12;\hskip5.4pt
                   \else\ifnum#5=5\hskip-9.2pt\vr1,29;\hskip8.8pt
                                   \fi\fi\fi\fi\fi

                   \ifnum#3=0\hskip8pt
                   \else\ifnum#3=1\hskip-5pt\hr13,5.4;
                   \else\ifnum#3=2\hskip-5.5pt\hr13.5,7.6;
                                  \hskip-13.5pt\hr13.5,3.2;
                   \else\ifnum#3=3\hskip-5.7pt\hr13.7,8.4;
                                  \hskip-13pt\hr13,5.4;
                                  \hskip-13.7pt\hr13.7,2.4;
                  \else\ifnum#3=7\hskip-5.7pt\hr13.7,8.4;
                                 \hskip-13.2pt\hr13.2,6.4;
                                 \hskip-13.2pt\hr13.2,4.4;
                                  \hskip-13.7pt\hr13.7,2.4;
                  \else\ifnum#3=8\hskip-5.7pt\hr13.7,8.4;
                                 \hskip-13.2pt\hr13.2,6.4;
                                 \hskip-13.2pt\hr13.2,4.4;
                                  \hskip-13.9pt$\langle$\hskip-8pt\hr13.7,2.4;
                   \else\ifnum#3=4\hskip-5.5pt\hr13.5,7.6;
                                  \hskip-13.5pt$\langle$\hskip-8.2pt
                                  \hr13.5,3.2;
                  \else\ifnum#3=5\hskip-5pt\hr30,5.4;
                   \else\ifnum#3=6\hskip-5.7pt\hr13.7,8.4;
                                  \hskip-13pt\hr13,5.4;
                                 \hskip-13.9pt$\langle$\hskip-8.0pt\hr13.7,2.4;
                                 \fi\fi\fi\fi\fi\fi\fi\fi\fi
                   }}
\def\blob#1#2#3#4#5#6#7{\hbox
{$\displaystyle\mathop{\blb#1#2#3#4#5 }_{#6}\sp{#7}$}}
\def\up#1#2{\dimen1=33pt\multiply\dimen1by#1
                  \hbox{\raise\dimen1\rlap{#2}}}
\def\uph#1#2{\dimen1=17.5pt\multiply\dimen1by#1
                  \hbox{\raise\dimen1\rlap{#2}}}
\def\dn#1#2{\dimen1=33pt\multiply\dimen1by#1
                   \hbox{\lower\dimen1\rlap{#2}}}
\def\dnh#1#2{\dimen1=17.5pt\multiply\dimen1by#1
                    \hbox{\lower\dimen1\rlap{#2}}}

\def\rlbl#1{\kern-8pt\raise3pt\hbox{$\scriptstyle #1$}}
\def\llbl#1{\raise3pt\llap{\hbox{$\scriptstyle #1$\kern-8pt}}}
\def\elbl#1{\kern3pt\lower4.5pt\hbox{$\scriptstyle #1$}}
\def\lelbl#1{\rlap{\hbox{\kern-9pt\raise2.5pt\hbox{{$\scriptstyle #1$}}}}}

\def\wht#1#2#3#4{\blob\circ#1#2#3#4{}{}}

\def\whtd#1#2#3#4#5{\blob\circ#1#2#3#4{#5}{}}
\def\blkd#1#2#3#4#5{\blob\bullet#1#2#3#4{#5}{}}
\def\whtu#1#2#3#4#5{\blob\circ#1#2#3#4{}{#5}}
\def\blku#1#2#3#4#5{\blob\bullet#1#2#3#4{}{#5}}
\def\whtr#1#2#3#4#5{\blob\circ#1#2#3#4{}{}\rlbl{#5}}

\def\whtl#1#2#3#4#5{\llbl{#5}\blob\circ#1#2#3#4{}{}}

\def\rwng{\hbox{$\vbox{\offinterlineskip{
  \hbox{\phantom{}\kern6pt{$\circ$}}\kern-2.5pt\hbox{$\Biggr/$}\kern-0.5pt
  \hbox{\phantom{}\kern-5pt$\circ$}\kern-3.0pt\hbox{$\Biggr\backslash$}
  \kern-1.5pt\hbox{\phantom{}\kern6pt{$\circ$}} }}$}}

\def\lwng{\hbox{$\vbox{\offinterlineskip{ \hbox{$\circ$}
  \kern-3.0pt\hbox{\phantom{}\kern6.0pt{$\Biggr\backslash$}}
  \kern-0.5pt\hbox{\phantom{}\kern11pt{$\circ$}}\kern-3.5pt
  \hbox{\phantom{}\kern5.0pt {$\Biggr/$}}\kern-1.0pt\hbox{$\circ$} }}$}}

\def\drwng#1#2#3{\hbox{$\vcenter{ \offinterlineskip{
  \hbox{\phantom{}\kern6pt{$\circ^{\elbl{#3}}$}}
  \kern-2.5pt\hbox{$\Biggr/$}\kern-0.5pt
  \hbox{\phantom{}\kern-5pt$\circ^{ \elbl{#1}}$}
  \kern-3.0pt\hbox{$\Biggr\backslash$}
  \kern-1.5pt\hbox{\phantom{}\kern6pt{$\circ^{\elbl{#2}}$}}  } }$}}

\def\dlwng#1#2#3{\hbox{$\vcenter{\offinterlineskip{ \hbox{$\lelbl{#1}\circ$}
  \kern-3.0pt\hbox{\phantom{}\kern6.0pt{$\Biggr\backslash$}}
  \kern-0.5pt\hbox{\phantom{}\kern11pt{$\lelbl{#2}\circ$}}\kern-3.5pt
  \hbox{\phantom{}\kern5.0pt {$\Biggr/$}}\kern-1.0pt\hbox{$\lelbl{#3}\circ$}}}$}
}

\def\rde#1#2#3{\raisebox{.5pt}{\hbox{\phantom{}\kern-4pt\hbox{$\vcenter
{\offinterlineskip\hbox{
               \raise 4.5pt\hbox{\vrule height0.4pt width13pt depth0pt}
                \kern-1pt\vbox{ \hbox{\drwng{#1}{#2}{#3}}} }}$  }} }}

\def\lde#1#2#3{\raisebox{.5pt}{\hbox{$\vcenter{\offinterlineskip  \hbox{
               \dlwng{#1}{#2}{#3}\kern-5.2pt\lower0.4pt\hbox{$\vcenter{\hrule
 width13pt}$}
               \kern-8pt\phantom{}   }}  $}}}

\def\ldet#1#2#3{\hbox{$\vcenter{\offinterlineskip  \hbox{
               \dlwng{#1}{#2}{#3}\kern14pt\lower0.4pt\hbox{$\vcenter{
          \hskip-20pt\hr13.5,5.7;\hskip-13.5pt\hr13.5,1.3;}$}
               \kern-25.8pt\phantom{}   }}  $}}

\def\rwngb{\hbox{$\vbox{\offinterlineskip{
  \hbox{\phantom{}\kern6pt{$\bullet$}}\kern-2.5pt\hbox{$\Biggr/$}\kern-0.5pt
  \hbox{\phantom{}\kern-5pt$\bullet$}\kern-3.0pt\hbox{$\Biggr\backslash$}
  \kern-1.5pt\hbox{\phantom{}\kern6pt{$\bullet$}} }}$}}

\def\lwngb{\hbox{$\vbox{\offinterlineskip{ \hbox{$\bullet$}
  \kern-3.0pt\hbox{\phantom{}\kern6.0pt{$\Biggr\backslash$}}
  \kern-0.5pt\hbox{\phantom{}\kern11pt{$\bullet$}}\kern-3.5pt
  \hbox{\phantom{}\kern5.0pt {$\Biggr/$}}\kern-1.0pt\hbox{$\bullet$} }}$}}

\def\dbrwng#1#2#3{\hbox{$\vcenter{ \offinterlineskip{
  \hbox{\phantom{}\kern6pt{$\bullet^{\elbl{#3}}$}}
  \kern-2.5pt\hbox{$\Biggr/$}\kern-0.5pt
  \hbox{\phantom{}\kern-5pt$\bullet^{ \elbl{#1}}$}
  \kern-3.0pt\hbox{$\Biggr\backslash$}
  \kern-1.5pt\hbox{\phantom{}\kern6pt{$\bullet^{\elbl{#2}}$}}  } }$}}

\def\dblwng#1#2#3{\hbox{$\vcenter{\offinterlineskip{ \hbox{$\lelbl{#1}\bullet$}
  \kern-3.0pt\hbox{\phantom{}\kern6.0pt{$\Biggr\backslash$}}
  \kern-0.5pt\hbox{\phantom{}\kern11pt{$\lelbl{#2}\bullet$}}\kern-3.5pt
  \hbox{\phantom{}\kern5.0pt
 {$\Biggr/$}}\kern-1.0pt\hbox{$\lelbl{#3}\bullet$}}}$} }

\def\rbde#1#2#3{\hbox{\phantom{}\kern-4pt\hbox{$\vcenter{\offinterlineskip
 \hbox{
               \raise 4.5pt\hbox{\vrule height0.4pt width13pt depth0pt}
                \kern-1pt\vbox{ \hbox{\dbrwng{#1}{#2}{#3}}} }}$  }}  }

\def\lbde#1#2#3{\hbox{$\vcenter{\offinterlineskip  \hbox{
               \dblwng{#1}{#2}{#3}\kern-4.2pt\lower0.4pt\hbox{$\vcenter{\hrule
 width13pt}$}
               \kern-8pt\phantom{}   }}  $}}

\def\ddgu#1.#2.{\dynk  \whtu0300{#1}\blku3000{#2}}
\def\ddgd#1.#2.{\dynk  \whtd0300{#1}\blkd3000{#2}}

\def\eddgiu#1.#2.#3.{\dynk \whtu0100{#1}\whtu1300{#2}\whtu6000{#3}}
\def\eddgiiu#1.#2.#3.{\dynk  \whtu0300{#1}\blku3100{#2}\blku1000{#3}}

\def\eddgid#1.#2.#3.{\dynk \whtd0100{#1}\whtd1300{#2}\whtd6000{#3}}
\def\eddgiid#1.#2.#3.{\dynk  \whtd0300{#1}\whtd6100{#2}\whtd1000{#3}}

\def\ddfu#1.#2.#3.#4.{\dynk
 \whtu0100{#1}\whtu1200{#2}\blku2100{#3}\blku1000{#4}}
\def\ddfd#1.#2.#3.#4.{\dynk
 \whtd0100{#1}\whtd1200{#2}\blkd2100{#3}\blkd1000{#4}}

\def\eddfiu#1.#2.#3.#4.#5.{\dynk
 \whtu0100{#1}\whtu1100{#2}\whtu1200{#3}\whtu4100{#4}\whtu1000{#5}}

\def\eddfid#1.#2.#3.#4.#5.{\dynk
 \whtd0100{#1}\whtd1100{#2}\whtd1200{#3}\whtd4100{#4}\whtd1000{#5}}
 
\def\eddfiiu#1.#2.#3.#4.#5.{\dynk
 \whtu0100{#1}\whtu1200{#2}\whtu4100{#3}\whtu1100{#4}\whtu1000{#5}}
\def\eddfiid#1.#2.#3.#4.#5.{\dynk
 \whtu0100{#1}\whtd1200{#2}\whtd4100{#3}\whtu1100{#4}\whtd1000{#5}}

\def\ddanu#1.#2.#3.#4.#5.{\dynk \whtu0100{#1}\whtu1100{#2}\whtu1100{#3}\cdots%
                           \whtu1100{#4}\whtu1000{#5}}
\def\ddanuf#1.#2.#3.#4.{\dynk \whtd0100{#1}\whtu1100{#2}\cdots%
                           \whtu1100{#3}\whtu1000{#4}}
\def\ddanuuf#1.#2.#3.#4.{\dynk \whtu0100{#1}\whtu1100{#2}\cdots%
                           \whtu1100{#3}\whtu1000{#4}}
\def\ddanufd#1.#2.#3.#4.{\dynk \whtd0100{#1}\whtu1100{#2}\cdots%
                           \whtu1100{#3}\whtr1005{#4}}
\def\ddandf#1.#2.#3.#4.{\dynk \whtu0100{#1}\whtd1100{#2}\cdots%
                           \whtd1100{#3}\whtd1000{#4}}
\def\ddanddf#1.#2.#3.#4.{\dynk \whtd0100{#1}\whtd1100{#2}\cdots%
                           \whtd1100{#3}\whtd1000{#4}}
\def\ddandfu#1.#2.#3.#4.{\dynk \whtu0100{#1}\whtd1100{#2}\cdots%
                           \whtd1100{#3}\whtr1050{#4}}
\def\ddcnds#1.#2.#3.#4.#5.#6{\dynk \whtd0200{#1}\whtd4100{#2}%
                          \whtd1100{#3}\cdots%
                           \whtd1100{#4}\whtd1400{#5}\whtd2000{#6}}

\def\eddanu#1.#2.#3.#4.#5.{\dynk \whtu0100{#1}\whtu1100{#2}%
                           \up1{\whtr0000{#3}}\cdots\whtu1100{#4}\whtu1000{#5}}
\def\eddand#1.#2.#3.#4.#5.{\dynk \whtd0100{#1}\whtd1100{#2}%
                           \up1{\whtr0000{#3}}\cdots\whtd1100{#4}\whtd1000{#5}}
\def\ddand#1.#2.#3.#4.#5.{\dynk \whtd0100{#1}\whtd1100{#2}\whtd1100{#3}\cdots%
                           \whtd1100{#4}\whtd1000{#5}}
\def\andfive#1.#2.#3.#4.#5.{\dynk \whtu0100{#1}\whtu1100{#2}\whtu1100{#3}%
                           \whtu1100{#4}\whtu1000{#5}}
\def\andthr#1.#2.#3.{\dynk \whtu0100{#1}\whtu1100{#2}\whtu1000{#3}}

\def\eddanid#1.#2.#3.#4.#5.{\dynk \whtd0200{#1}\whtd4100{#2}%
                           \whtd1100{#3}\cdots\whtd1200{#4}\whtd4000{#5}}
\def\eddanidr#1.#2.#3.#4.#5.{\dynk \whtd0200{#1}\whtd4100{#2}%
                           \cdots\whtd1100{#3}\whtd1200{#4}\whtd4000{#5}}
\def\eddaniu#1.#2.#3.#4.#5.{\dynk \whtu0200{#1}\whtu2100{#2}%
                           \whtu1100{#3}\cdots\whtu1200{#4}\blku2000{#5}}

\def\eddaniid#1.#2.#3.#4.#5.#6.{\hbox{$\vcenter{\hbox
         {\dynk\hbox{$ \lde{#1}{#2}{#3}\whtd1100{#4}\cdots%
          \whtd1400{#5}\whtd2000{#6} $}} }$}}

\def\eddaniiu#1.#2.#3.#4.#5.#6.{\hbox{$\vcenter{\hbox
         {\dynk\hbox{$ \lbde{#1}{#2}{#3}\blku1100{#4}\cdots%
          \blku1200{#5}\whtu2000{#6} $}} }$}}

\def\eddaiii#1.#2.#3.{\dynk\whtd0400{#1}\whtd2200{#2}\whtd4000{#3}}
\def\eddaiiif#1.#2.#3.#4.{\dynk\whtd0400{#1}\whtd2100{#2}%
                    \whtd1200{#3}\whtd4000{#4}}
\def\eddciii#1.#2.#3.{\dynk\whtd0200{#1}\whtd4400{#2}\whtd2000{#3}}

\def\ddbnu#1.#2.#3.#4.#5.{\dynk \whtu0100{#1}\whtu1100{#2}\whtu1100{#3}\cdots%
                           \whtu1200{#4}\blku2000{#5}}
\def\ddbnd#1.#2.#3.#4.#5.{\dynk \whtd0100{#1}\whtd1100{#2}\whtd1100{#3}\cdots%
                           \whtd1200{#4}\blkd2000{#5}}

\def\eddbnu#1.#2.#3.#4.#5.#6.{\dynk \lde{#1}{#2}{#3}\whtu1100{#4}\cdots%
                           \whtu1200{#5}\blku2000{#6}}

\def\eddbnd#1.#2.#3.#4.#5.#6.{\dynk \lde{#1}{#2}{#3}\whtd1100{#4}\cdots%
                           \whtd1200{#5}\whtd4000{#6}}
\def\eddbndt#1.#2.#3.#4.{\dynk \ldet{#1}{#2}{#3}\hskip-444pt\whtr4000{#4}}

\def\ddcnu#1.#2.#3.#4.#5.{\dynk \blku0100{#1}\blku1100{#2}\blku1100{#3}\cdots%
                          \blku1200{#4}\whtu2000{#5}}
\def\ddcnd#1.#2.#3.#4.#5.{\dynk \blkd0100{#1}\blkd1100{#2}\blkd1100{#3}\cdots%
                           \blkd1200{#4}\whtd2000{#5}}

\def\eddcnu#1.#2.#3.#4.#5.#6.{\dynk \whtu0200{#1}\blku2100{#2}\blku1100{#3}%
       \blku1100{#4}\cdots\blku1200{#5}\whtu2000{#6}}

\def\ncddlr#1.#2.#3.#4.#5.{\dynk \whtu0101{#1}\whtu1100{#2}\whtu1100{#3}\cdots%
                           \whtu1100{#4}\whtu1001{#5}}
\def\ncdulr#1.#2.#3.#4.#5.{\dynk \whtd0110{#1}\whtd1100{#2}\whtd1100{#3}\cdots%
                           \whtd1100{#4}\whtd1010{#5}}
\def\ncddrd#1.#2.#3.#4.#5.{\dynk \whtd0100{#1}\whtu1100{#2}\whtu1100{#3}\cdots%
                           \whtu1100{#4}\whtu1005{#5}}
\def\ncddru#1.#2.#3.#4.#5.{\dynk \whtu0100{#1}\whtd1100{#2}\whtd1100{#3}\cdots%
                           \whtd1100{#4}\whtd1050{#5}}
\def\ncddrdu#1.#2.#3.#4.#5.{\dynk \whtu0100{#1}\whtu1100{#2}\whtu1100{#3}%
                         \cdots\whtu1100{#4}\whtu1005{#5}}
\def\ncddrud#1.#2.#3.#4.#5.{\dynk \whtd0100{#1}\whtd1100{#2}\whtd1100{#3}%
                            \cdots\whtd1100{#4}\whtd1050{#5}}
\def\ncddld#1.#2.#3.#4.#5.{\dynk \whtl0105{#1}\whtu1100{#2}\whtu1100{#3}\cdots%
                           \whtu1100{#4}\whtd1000{#5}}
\def\ncddlu#1.#2.#3.#4.#5.{\dynk \whtl0150{#1}\whtd1100{#2}\whtd1100{#3}\cdots%
                           \whtd1100{#4}\whtu1000{#5}}
\def\ncanur#1.#2.#3.#4.#5.{\dynk \whtu0101{#1}\whtu1100{#2}\whtu1100{#3}\cdots%
                           \whtu1100{#4}\whtd1010{#5}}
\def\ncandr#1.#2.#3.#4.#5.{\dynk \whtd0110{#1}\whtd1100{#2}\whtd1100{#3}\cdots%
                           \whtd1100{#4}\whtu1001{#5}}

\def\eddcnd#1.#2.#3.#4.#5.{\dynk \whtd0200{#1}\whtd4100{#2}\whtd1100{#3}
       \cdots \whtd1400{#4}\whtd2000{#5}}

\def\dddnu#1.#2.#3.#4.#5.#6.{\hbox{$\vcenter{\hbox
         {\dynk\hbox{$ \whtu0100{#1}\whtu1100{#2}\cdots%
          \whtu1100{#3}\rde{#4}{#5}{#6} $}}  }$}}
\def\dddnd#1.#2.#3.#4.#5.#6.{\hbox{$\vcenter{\hbox
         {\dynk\hbox{$ \whtd0100{#1}\whtd1100{#2}\cdots%
          \whtd1100{#3}\rde{#4}{#5}{#6} $}} }$}}
\def\dddiv#1.#2.#3.#4.{\hbox{$\vcenter{\hbox
         {\dynk\hbox{$ \whtu0100{#1}\rde{#2}{#3}{#4}
              $}}  }$}}
\def\edddiv#1.#2.#3.#4.{\hbox{$\vcenter{\hbox{\dynk\hbox{$\whtl0100{#1}
\up1{\whtl0001{#2}}\dn1{\whtl0010{#4}}\wht1111\whtr1000{#3} $}}}$}}

\def\edddnu#1.#2.#3.#4.#5.#6.#7.#8.{\hbox{$\vcenter{\hbox
         {\dynk\hbox{$ \lde{#1}{#2}{#3}\whtu1100{#4}\cdots%
          \whtu1100{#5}\rde{#6}{#7}{#8} $}}  }$}}
\def\edddnd#1.#2.#3.#4.#5.#6.#7.#8.{\hbox{$\vcenter{\hbox
         {\dynk\hbox{$ \lde{#1}{#2}{#3}\whtd1100{#4}\cdots%
          \whtd1100{#5}\rde{#6}{#7}{#8} $}} }$}}
\def\edddndf#1.#2.#3.#4.#5.#6.{\hbox{$\vcenter{\hbox
         {\dynk\hbox{$ \lde{#1}{#2}{#3}\rde{#4}{#5}{#6} $}} }$}}

\def\edddnds#1.#2.#3.#4.#5.#6.#7.#8.#9.{\hbox{$\vcenter{\hbox
      {\dynk\hbox{$ \lde{#1}{#2}{#3}\whtd1100{#4}\cdot\cdot\whtd1100{#5}\cdot%
      \cdot\whtd1100{#6}\rde{#7}{#8}{#9} $}} }$}}
\def\eddanod#1.#2.#3.#4.#5.#6.{\hbox{$\vcenter{\hbox
         {\dynk\hbox{$ \whtd0200{#1}\whtd4100{#2}\cdots%
          \whtd1100{#3}\rde{#4}{#5}{#6} $}} }$}}
		  
\def\edddniid#1.#2.#3.#4.#5.{\hbox{$\vcenter{\hbox
         {\dynk\hbox{$ \whtd0400{#1}\whtd2100{#2}\whtd1100{#3}\cdots%
          \whtd1200{#4}\whtd4000{#5} $}} }$}}

\def\edddniiu#1.#2.#3.#4.#5.{\hbox{$\vcenter{\hbox
         {\dynk\hbox{$ \blku0200{#1}\whtu2100{#2}\whtu1100{#3}\cdots%
          \whtu1200{#4}\blku2000{#5} $}} }$}}

\def\ddei#1.#2.#3.#4.#5.#6.{\hbox{$\vcenter{\hbox
       {\dynk \whtd0100{#1}\whtd1100{#3}%
       \up1{\whtr0001{#2}}\whtd1110{#4}\whtd1100{#5}\whtd1000{#6}} }$}}

\def\eddei#1.#2.#3.#4.#5.#6.#7.{\hbox{$\vcenter{\hbox
       {\dynk \whtu0100{#1}\whtu1100{#3}%
       \up1{\whtr0011{#2}}\up2{\whtr0001{#7}}\whtd1110{#4}\whtu1100{#5}%
       \whtu1000{#6}} }$}}

\def\ncdddt#1.#2.{\dynk\whtu0400{#1}\whtu2001{#2}}
\def\ncandrt#1.#2.{\dynk\whtd0110{#1}\whtu1001{#2}}
\def\ncanurt#1.#2.{\dynk\whtu0101{#1}\whtd1010{#2}}
\def\ncddet#1.#2.{\dynk\whtu0400{#1}\whtu2000{#2}}
\def\ncddut#1.#2.{\dynk\whtd0400{#1}\whtd2010{#2}}
\def\ncdduot#1.#2.{\dynk\whtd0210{#1}\whtd4000{#2}}
\def\ncddct#1.#2.{\hbox{\dynk\whtu0200{\rotatebox{45}{$\scriptstyle#1$}}%
                    \whtu4000{\rotatebox{45}{$\scriptstyle#2$}}}}
\def\ncddcot#1.#2.{\dynk\whtd0400{\rotatebox{45}{$\scriptstyle#1$}}%
                    \whtd2000{\rotatebox{45}{$\scriptstyle#2$}}}
\def\ncddcst#1.#2.{\dynk\whtu0400{\rotatebox{135}{$\scriptstyle#1$}}%
                    \whtu2000{\rotatebox{135}{$\scriptstyle#2$}}}
           
\def\rronit#1.{\rotatebox{315}
       {\dynk\whtr5005{\rotatebox{45}{$\scriptstyle #1$}}}}
\def\laronit#1.#2.{\rotatebox{315}{\ncddct#1.#2.}}
\def\raronit#1.#2.{\rotatebox{315}{\ncddcot#1.#2.}}
\def\rarsnit#1.#2.{\rotatebox{225}{\ncddcst#1.#2.}}

\def\ncddd#1.#2.#3.#4.#5.{\dynk\whtu0400{#1}\whtu2100{#2}\whtu1100{#3}%
           \cdots\whtu1100{#4}\whtu1001{#5}}
\def\ncdde#1.#2.#3.#4.#5.{\dynk\whtu0400{#1}\whtu2100{#2}\whtu1100{#3}%
           \cdots\whtu1100{#4}\whtu1000{#5}}
\def\ncdded#1.#2.#3.#4.#5.{\dynk\whtl0400{#1}\whtd2100{#2}\whtd1100{#3}%
           \cdots\whtd1100{#4}\whtd1000{#5}}
\def\ncddeo#1.#2.#3.#4.#5.{\dynk\whtd0100{#1}\whtd1100{#2}\whtd1100{#3}%
           \cdots\whtd1200{#4}\whtd4000{#5}}
\def\ncddeof#1.#2.#3.#4.{\dynk\whtd0100{#1}\whtd1100{#2}%
           \cdots\whtd1200{#3}\whtr4000{#4}}
\def\ncddu#1.#2.#3.#4.#5.{\dynk\whtd0400{#1}\whtd2100{#2}\whtd1100{#3}%
           \cdots\whtd1100{#4}\whtd1010{#5}}
\def\ncdduo#1.#2.#3.#4.#5.{\dynk\whtd0110{#1}\whtd1100{#2}\whtd1100{#3}%
           \cdots\whtd1200{#4}\whtd4000{#5}}
\def\ncddc#1.#2.#3.#4.#5.{\dynk\whtu0200{#1}\whtu4100{#2}\whtu1100{#3}%
           \cdots\whtu1100{#4}\whtu1000{#5}}
\def\ncdddc#1.#2.#3.#4.#5.{\dynk\whtd0200{#1}\whtd4100{#2}\whtd1100{#3}%
           \cdots\whtd1100{#4}\whtd1000{#5}}
\def\ncddcu#1.#2.#3.#4.#5.{\dynk\whtl0200{#1}\whtd4100{#2}\whtd1100{#3}%
           \cdots\whtd1100{#4}\whtr1050{#5}}
\def\ncddcd#1.#2.#3.#4.#5.{\dynk\whtu0200{#1}\whtu4100{#2}\whtu1100{#3}%
           \cdots\whtu1100{#4}\whtu1005{#5}}
\def\ncddco#1.#2.#3.#4.#5.{\dynk\whtd0100{#1}\whtd1100{#2}\cdots%
            \whtd1100{#3}\whtd1400{#4}\whtd2000{#5}}
\def\ncdfr#1.#2.#3.#4.#5.#6.{\ncddc#1.#2.#3.#4.#5.
           \hskip-42.5pt\dynk\rotatebox{315}
          {\whtu5005{\rotatebox{45}{$\scriptstyle #6$}}}}
\def\ncdfrd#1.#2.#3.#4.#5.#6.{\ncdde#1.#2.#3.#4.#5.
           \hskip-42.5pt\dynk\rotatebox{315}
          {\whtr5005{\rotatebox{45}{$\scriptstyle #6$}}}}
\def\ncdfrdc#1.#2.#3.#4.#5.#6.{\ncddc#1.#2.#3.#4.#5.
           \hskip-42.5pt\dynk\rotatebox{315}
          {\whtr5005{\rotatebox{45}{$\scriptstyle #6$}}}}
\def\ncdfrdl#1.#2.#3.#4.#5.#6.{\ncddlu#1.#2.#3.#4.#5.
           \hskip-42.5pt\dynk\rotatebox{315}
          {\whtr5005{\rotatebox{45}{$\scriptstyle #6$}}}}

\def\ncdfl#1.#2.#3.#4.#5.#6.{\hbox{\lronit#1.\hskip-39.5pt
                   \raisebox{13pt}{$\ddand#2.#3.#4.#5.#6.$}}}
\def\ncdfal#1.#2.#3.#4.#5.{\hbox{\lronit#1.\hskip-39.5pt
                   \raisebox{13pt}{$\ddandf#2.#3.#4.#5.$}}}
\def\ncdfalu#1.#2.#3.#4.#5.{\hbox{\lronit#1.\hskip-39.5pt
                   \raisebox{13pt}{$\ddandfu#2.#3.#4.#5.$}}}

\def\ncdfar#1.#2.#3.#4.#5.{\ddanuf#1.#2.#3.#4.
  \hskip-42.5pt\dynk\rotatebox{315}{\whtr5005{\rotatebox{45}
           {$\scriptstyle #5$}}}}
\def\ncdfaur#1.#2.#3.#4.#5.{\ddanuuf#1.#2.#3.#4.
  \hskip-42.5pt\dynk\rotatebox{315}{\whtu5005{\rotatebox{45}
           {$\scriptstyle #5$}}}}

\def\lronit#1.{\rotatebox{315}
       {\dynk\whtl0550{\rotatebox{45}{$\scriptstyle #1$}}}}
\def\daone#1.#2.{\dynk\whtd0400{#1}\whtd4000{#2}}

\def\datwot#1.#2.{\dynk\whtu0700{#1}\whtu8000{#2}}

\def\datwon#1.#2.#3.#4.#5.#6.{\dynk \whtd0200{#1}\whtd4100{#2}\whtd1100{#3}%
          \whtd1100{#4} \cdots\whtd1200{#5}\whtd4000{#6}}
\def\datwono#1.#2.#3.#4.#5.#6.{\dynk \whtd0400{#1}\whtd2100{#2}\whtd1100{#3}%
         \cdots \whtd1100{#4}\whtd1400{#5}\whtd2000{#6}}
\def\datwonl#1.#2.#3.#4.#5.#6.{\dynk \whtd0200{#1}\whtd4100{#2}\whtd1100{#3}%
         \cdots \whtd1100{#4} \whtd1200{#5}\whtd4000{#6}}

\def\ddeii#1.#2.#3.#4.#5.#6.#7.{\hbox{$\vcenter{\hbox
       {\dynk \whtd0100{#1}\whtd1100{#3}%
       \up1{\whtr0001{#2}}\whtd1110{#4}\whtd1100{#5}\whtd1100{#6}%
       \whtd1000{#7}} }$}}

\def\eddeii#1.#2.#3.#4.#5.#6.#7.#8.{\hbox{$\vcenter{\hbox
       {\dynk \whtu0100{#8}\whtu1100{#1}\whtu1100{#3}%
       \up1{\whtr0001{#2}}\whtd1110{#4}\whtu1100{#5}\whtu1100{#6}%
       \whtu1000{#7}} }$}}

\def\ddeiii#1.#2.#3.#4.#5.#6.#7.#8.{\hbox{$\vcenter{\hbox
       {\dynk \whtd0100{#1}\whtd1100{#3}%
       \up1{\whtr0001{#2}}\whtd1110{#4}\whtd1100{#5}\whtd1100{#6}%
       \whtd1100{#7}\whtd1000{#8}} }$}}

\def\eddeiii#1.#2.#3.#4.#5.#6.#7.#8.#9.{\hbox{$\vcenter{\hbox
       {\dynk \whtd0100{#1}\whtd1100{#3}%
       \up1{\whtr0001{#2}}\whtd1110{#4}\whtd1100{#5}\whtd1100{#6}%
       \whtd1100{#7}\whtd1100{#8}\whtd1000{#9}} }$}}

\begin{document}
\newfont{\elevenmib}{cmmib10 scaled\magstep1}%
\newfont{\cmssbx}{cmssbx10 scaled\magstep3}
\newcommand{\preprint}{
            \begin{flushleft} 
			\elevenmib Yukawa\, Institute\, Kyoto\\
            \end{flushleft}\vspace{-1.3cm}
            \begin{flushright}\normalsize  \sf
            YITP-98-60\\
            KUCP-0121\\ 
			{\tt hep-th/9809068} \\ September 1998
            \end{flushright}}
\newcommand{\Title}[1]{{\baselineskip=26pt \begin{center} 
            \Large   \bf #1 \\ \ \\ \end{center}}}
\newcommand{\Author}{\begin{center}\large \bf
            A.\, J.\, Bordner$^a$, \ \
            R.\, Sasaki$^a$ and K. Takasaki$^b$\end{center}}
\newcommand{\Address}{\begin{center} \it 
            $^a$ Yukawa Institute for Theoretical Physics, Kyoto
            University,\\ Kyoto 606-8502, Japan \\
            $^b$  Department of Fundamental Sciences,
            Faculty of Integrated Human Studies,\\ Kyoto University,
            Kyoto 606-8501,Japan
		    \end{center}}
\newcommand{\Accepted}[1]{\begin{center}{\large \sf #1}\\
            \vspace{1mm}{\small \sf Accepted for Publication}
            \end{center}}
\baselineskip=20pt

\preprint
\thispagestyle{empty}
\bigskip
\bigskip
\bigskip
\Title{Calogero-Moser Models II: Symmetries and Foldings}
\Author

\Address
\vspace{2cm}

\begin{abstract}%
\noindent
Universal Lax pairs (the root type and the minimal type)
 are presented for Calogero-Moser models based on simply laced
root systems, including \(E_8\).
They exist with and without spectral parameter and they work
for all of the four choices of  
potentials: the rational, trigonometric, hyperbolic and elliptic.
For the elliptic potential, the discrete symmetries
of the simply laced models, originating from the automorphism
of the extended Dynkin diagrams, are combined
with the periodicity of the potential to derive a
class of Calogero-Moser models known as the
`twisted non-simply laced models'.
For untwisted non-simply laced models, two kinds of
root type Lax pairs (based on long roots and short roots)
are derived which contain independent coupling constants
for the long and short roots.
The \(BC_n\) model contains three independent couplings,
for the long, middle and short roots.
The \(G_2\) model based on long roots
exhibits a new feature which deserves further study.
\\ \\
\end{abstract}

\Accepted{Progress of Theoretical Physics, {\bf 101} (1999)}
\newpage

\section{Introduction}
\setcounter{equation}{0}
In a previous paper \cite{bcs} a new and universal formulation of
Lax pairs of Calogero-Moser models based on
simply laced root systems was presented.
This paper is devoted to further developments and refinements
of the Lax pairs \cite{OP1,Kri,Ino,DHPh} and the Calogero-Moser models 
themselves with an emphasis on the symmetries of the simply laced as well
as the twisted and untwisted non-simply laced models.
The Calogero-Moser models \cite{CalMo} are a collection 
of completely
integrable one-dimensional 
dynamical systems characterised by root systems and a choice of
four long-range  interaction potentials:
 (i) $1/L^2$, (ii) $1/\sin^2L$, (ii) $1/\sinh^2L$
and (iv) $\wp(L)$, in which $L$ is the inter-particle ``distance''.

Besides various direct applications of the models to lower dimensional
physics ranging from solid state  to particle physics \cite{ss},
elliptic Calogero-Moser models are attracting attention owing
to their connection with (supersymmetric) gauge
theory,  classical soliton dynamics \cite{Kri}, 
Toda theories and infinite dimensional 
algebras.
The Seiberg-Witten curve and 
differential and ${\cal N}=2$ supersymmetric gauge theory 
are analysed in terms of elliptic Calogero-Moser models
with the same Lie algebra \cite{SeiWit}-\cite{DHPh1}.
The untwisted and twisted Calogero-Moser models are known
to reduce to Toda models in a certain limit  \cite{Ino,DHPh}.
The affine algebras acting on Toda models are relatively well understood.
This fosters an expectation that the elliptic Calogero-Moser models
(with the Lie algebraic aspects from the root system and the
toroidal aspects from the potential) open a way to a
greater symmetry algebra than the
affine algebras \cite{GN}.

In this paper we address  the problem of the
symmetries of the Calogero-Moser models and the associated Lax pairs,
in particular, the amalgamation of the Lie algebraic aspects
originating from the root structure and the toroidal aspects
from the elliptic potential.
As a first step we present the universal Lax pairs with 
and without spectral parameter for all  four choices of
potential  for Calogero-Moser models based on simply laced
root systems. There are two types of universal Lax pairs,
the root type and  the minimal type \cite{bcs}.
The root type Lax pair is represented on the set of roots itself.
It is intrinsic to the root system and it applies to all of
the models based on  root systems, including \(E_8\), for
which construction of a Lax pair had been a mystery for more
than twenty years. The minimal type Lax pair is represented
on the set of weights belonging to a minimal representation
\cite{bcs,GoOl}. Every Lie algebra, except for \(E_8\), has
at least one minimal representation. The minimal type Lax
pair provides  a unified description of all known examples
of  Calogero-Moser Lax pairs
and adds more \cite{bcs,OP1,Kri,Ino,DHPh}.

As a second step, we uncover a discrete symmetry of 
elliptic Calogero-Moser models based on simply laced root systems.
All simply laced root systems, except for \(E_8\), have a
symmetry under the automorphism(s) of the Dynkin diagram or
its  extended version.
By combining the symmetry under the automorphism with the
periodicity of the elliptic potential, a non-trivial
discrete symmetry of the models is obtained.
New integrable dynamical systems can be derived from
the elliptic Calogero-Moser models by restricting the 
dynamical variables to the invariant subspace of the 
discrete symmetry. This process is known as reduction or folding.
It is an important and useful tool in Toda lattices and field theories
\cite{folding,bcds}, another class of integrable models
based on root systems.
In the present case we obtain so-called twisted
non-simply laced Calogero-Moser models.

The untwisted non-simply laced models can also be obtained
by folding the simply laced models \cite{bcs}.
In the reduced models, however, the coupling 
constants for the long and short roots
have a fixed ratio, since the simply laced models 
have only one coupling.
In order to exhibit the fuller symmetry
of the untwisted non-simply laced models, root type Lax pairs 
with independent coupling constants are constructed as a third step.
There are two kinds of root type Lax pairs for non-simply laced models,
 one based on long roots and another on short
roots. Both are  straightforward generalisations
of the root type Lax pair for simply laced systems,
except for the \(G_2\) case based on long roots.
This case requires a new set of functions in the Lax
pair. A simple example of the new set of functions is given. The
\(BC_n\) model contains three independent couplings, for the
long, middle and short roots.

This paper is organised as follows.
In section two we present the universal Lax pairs with 
and without spectral parameter for all  four
choices of potential  for Calogero-Moser models based
on simply laced root systems. In section three certain
discrete symmetries of Calogero-Moser models  based on
simply laced root systems are introduced. Twisted
non-simply laced Calogero-Moser models are derived by 
folding with respect to this symmetry. In section four two
kinds of root type Lax pairs with independent coupling
constants, the one based on  long roots and the other
on short roots, are constructed for all of the untwisted
non-simply laced models. The \(BC_n\) model has three
independent couplings. Section five is devoted to
summary and comments.

\section{Universal Lax Pairs for Calogero-Moser Models
Based on  Simply Laced Algebras}
\setcounter{equation}{0}
In order to set the stage and introduce notation, let us recapitulate
our previous results of the universal Lax pairs for  the Calogero-Moser 
models
based on a  {\em simply laced} root system
\(\Delta\). For the elliptic potential the
universal Lax pairs without spectral  parameter 
were reported in a previous paper
\cite{bcs}. Here we include those with a spectral
parameter.

\bigskip
The basic ingredient of the model is a root system \(\Delta\)
associated with semi-simple and {\em simply laced}
Lie algebra 
${\mathfrak g}$ with rank $r$.
The roots \(\alpha,\beta,\gamma,\ldots\/\) are real $r$ 
dimensional vectors
and are normalised, without loss of generality, to 2:
\begin{equation}
   	\Delta=\{\alpha,\beta,\gamma,\ldots\}, \quad \alpha\in 
    {\mathbb R}^r,\quad
   	\alpha^2=\alpha\cdot\alpha=2,\quad \forall\alpha\in\Delta.
   	\label{eq:setroots}
\end{equation}
We denote by $Dim$ the total number of roots of $\Delta$. 

The dynamical variables are canonical coordinates $\{q^j\}$ and their 
canonical conjugate momenta $\{p_j\}$ with the Poisson brackets:
\begin{equation}
   	q^1,\ldots,q^r, \quad
    p_{1},\ldots,p_{r}, \quad
    \{q^j,p_{k}\}=\delta_{j,k},\quad \{q^j,q^{k}\}=\{p_{j},p_{k}\}=0.
   	\label{eq:poisson}
\end{equation}
In most cases we denote them by $r$ dimensional vectors $q$ and $p$
\footnote{
For  $A_r$ models, it is customary to introduce one more degree of 
freedom,
$q^{r+1}$ and $p_{r+1}$ and embed 
all of the roots in ${\mathbb R}^{r+1}$.},
\begin{displaymath}
   	q=(q^1,\ldots,q^r)\in {\mathbb R}^r,\quad
   	p=(p_1,\ldots,p_r)\in {\mathbb R}^r,\quad
\end{displaymath}
so that the scalar products of $q$ and $p$ with the roots 
$\alpha\cdot q$, $p\cdot\beta$, etc. can be defined.
The Hamiltonian is given by ($g$ is a real coupling constant)
\begin{equation}
   	{\cal H}={1\over2}p^2-{g^2\over2}\sum_{\alpha\in\Delta}
   	x(\alpha\cdot q)x(-\alpha\cdot q),
   	\label{eq:hamiltonian}
\end{equation}
in which \(x(t)\) is given
(\ref{eq:functions}--\ref{eq:functionsell6}) for various choices
of  potentials.

\bigskip
As is well known, with the help of a Lax pair,
\(L\) and \(M\),  which  expresses the
the canonical equation of motion derived from 
the  Hamiltonian 
(\ref{eq:hamiltonian}) in an equivalent matrix form:
\begin{equation}
   	\dot{L}={d\over{dt}}L=[L,M],
   	\label{eq:laxeq0}
\end{equation}
a sufficient number of conserved quantities can be
obtained by the trace:
\begin{equation}
   	{d\over{dt}}Tr(L^k)=0,\quad k=1,\ldots,.
\end{equation}
Two types of universal Lax pairs, the root type and the minimal type,
were constructed.
The matrices used in the root type Lax pair bear a resemblance to
the adjoint  representation of the associated Lie algebra, and
they exist for all models. Thus the root type Lax pair provides a
universal tool for proving the integrability of Calogero-Moser
models. The `minimal' types provide a unified description of all
known  examples of Calogero-Moser Lax pairs.
They are based on the set of weights of minimal representations
of the associated Lie algebras.
The important guiding principle for deriving these Lax pairs is the
Weyl invariance of the set of roots \(\Delta\) and of
the Hamiltonian. For details, see our previous paper \cite{bcs}.

\subsection{Root type Lax pair}
The detailed structure of the simply laced root system
\(\Delta\) is very different from one type of algebra to
another, which is hardly universal.
One universal feature is the root difference pattern,
i.e., which multiples of roots appear in the
difference of two roots:
\begin{equation}
   \mbox{Simply laced root system}:\qquad \mbox{ root}
   - \mbox{ root}=\left\{
   \begin{array}{l}
      \mbox{root}\\
      2\times \mbox{root}\\
      \mbox{non-root}
   \end{array}
   \right.
   \label{rootscheme}
\end{equation}
To be more specific, there can be no terms like
\(3\times\) root, etc. in the right hand side.
This then determines
 the root type Lax pair for simply laced root systems (we
choose
\(L\) to  be hermitian and \(M\) anti-hermitian):
\begin{eqnarray}
   	L(q,p,\xi) & = & p\cdot H + X + X_{d},\quad \nonumber\\
   	M(q,\xi) & = & D+Y+Y_{d},
   	\label{eq:genLaxform}
\end{eqnarray}
in which \(\xi\) is a spectral 	parameter, 
relevant only for the  elliptic potential.
Here $L$, $H$, $X$, $X_{d}$, $D$, $Y$ and $Y_{d}$ are $Dim\times 
Dim$ matrices whose indices are labelled by the roots themselves, 
 denoted here by 
$\alpha$,
$\beta$, $\gamma$, $\eta$ and $\kappa$. 
$H$ and $D$ are diagonal:
\begin{equation}
   	H_{\beta \gamma}=\beta \delta_{\beta, \gamma},\quad
   	D_{\beta \gamma}= \delta_{\beta, \gamma}D_{\beta},\quad
   	D_{\beta}=-ig\left(z(\beta\cdot q)+\sum_{\kappa\in\Delta,\  
   	\kappa\cdot\beta=1}z(\kappa\cdot q)\right).
   	\label{eq:HD}
\end{equation}
$X$ and $Y$ correspond to the first line of
(\ref{rootscheme}):
\begin{equation}
   	X=ig\sum_{\alpha\in\Delta}x(\alpha\cdot
    q, \xi)E(\alpha),\quad
   	Y=ig\sum_{\alpha\in\Delta}y(\alpha\cdot
    q, \xi)E(\alpha),\quad
   	E(\alpha)_{\beta \gamma}=\delta_{\beta-\gamma,\alpha}.
   	\label{eq:XYdef}
\end{equation}
$X_d$ and $Y_d$ are associated with the `double root'
in the second line of (\ref{rootscheme}):
\begin{equation}
   	X_d=2ig\sum_{\alpha\in\Delta}
    x_{d}(\alpha\cdot q, \xi)E_{d}(\alpha),\quad
   	Y_d=ig\sum_{\alpha\in\Delta}
    y_{d}(\alpha\cdot q, \xi)E_{d}(\alpha),\quad
   	E_{d}(\alpha)_{\beta \gamma}=\delta_{\beta-\gamma,2\alpha}.
   	\label{eq:XYrdef}
\end{equation}
The matrix $E(\alpha)$ ($E_{d}(\alpha)$) might be called a (double)
root discriminator. It takes the value one only when the difference
of the two indices is equal to  (twice) the root $\alpha$.
They correspond to the first and the second line of
(\ref{rootscheme}), respectively. Later in section 4 we
will encounter a triple root discriminator 
corresponding to the `\(3\times\) short root' part of
(\ref{g2longlong}). The functions $x,y,z$
($x_d,y_d,z_d$) depend on the choice  of the
inter-particle potential. For  the rational potential,
$1/L^2$,
 they are:
\begin{equation}
   	x(t)=x_{d}(t)={1\over t},\quad y(t)=y_{d}(t)=-{1\over{t^2}},
   	\quad z(t)=z_{d}(t)=-{1\over{t^2}}.
   	\label{eq:functions}
\end{equation}
For  the trigonometric potential, $1/\sin^2L$,  they
are:
\begin{equation}
   		x(t)=x_{d}(t)=a\cot at,\quad y(t)=y_{d}(t)=
     -{a^{2}\over{\sin^2 at}},
   			\quad z(t)=z_{d}(t)=-{a^{2}\over{\sin^2 at}},\quad a: const.
   	\label{eq:functionstri}
\end{equation}	
For  the hyperbolic potential, $1/\sinh^2L$,  they are:
\begin{equation}
   		x(t)=x_{d}(t)=a\coth at,\quad y(t)=y_{d}(t)=
     -{a^{2}\over{\sinh^2 at}},
   			\quad z(t)=z_{d}(t)=-{a^{2}\over{\sinh^2 at}}.
   	\label{eq:functionshyp}
\end{equation}	
For  the elliptic potential, $ \wp(L)$,  the functions
$x$ and $x_{d}$ generally differ. There are several
choices of the functions. They are related to each
other by a modular transformation. A first choice is
\footnote{We denote  the
fundamental periods of the Weierstrass' functions  by
\(\{2\omega_1,2\omega_3\}\) and \(e_j=\wp(\omega_j)\),
\(j=1,.,3\).}, 
\begin{eqnarray}
   		x(t)&=&{c\over2}\left[ {{1+k\, {\rm
sn}^2(ct/2,k)}
   		\over {\rm sn}(ct/2,k)}-
     i{{(1+k)(1-k\, {\rm sn}^2(ct/2,k))}\over{{\rm
cn}(ct/2,k)\,
     {\rm dn}(ct/2,k)}}\right],\nonumber\\
      y(t)&=&x^\prime(t),
   			\quad z(t)=-\wp(t),\quad
c=\sqrt{e_{1}-e_{3}},
   	\label{eq:functionsell1}
\end{eqnarray}
and
\begin{equation}
   		x_{d}(t)={c\over{{\rm sn}(ct,k)}},\quad 
   		y_{d}(t)=-c^{2}{{\rm cn}(ct,k)\,{\rm dn}(ct,k)
     \over{{\rm sn}^2(ct,k)}},
   			\quad z_{d}(t)=-\wp(t),
   	\label{eq:functionsell2}
\end{equation}
in which $k$ is the modulus of the elliptic function.
This set of functions
\footnote{The detailed properties of the functions in
the elliptic potential cases
 will be discussed elsewhere.} is obtained by setting 
\(\xi=\omega_3\)  in the
spectral parameter dependent functions (\ref{simplesol2}) for
\(j=3\).
\\
A second choice is
\begin{eqnarray}
   		x(t)&=&{c\over2}\left[ {{{\rm cn}^{2}(ct/2,k)-
     k^{\prime}{\rm sn}^2(ct/2,k)}
   		\over {\rm sn}(ct/2,k)\,{\rm cn}(ct/2,k)}+
     (1+k^{\prime}){{\rm
cn}^{2}(ct/2,k)+k^{\prime}{\rm 
     sn}^2(ct/2,k)\over{{\rm
dn}(ct/2,k)}}\right],\nonumber\\
     y(t)&=&x^\prime(t),
   			\quad z(t)=-\wp(t),
   	\label{eq:functionsell3}
\end{eqnarray}
and
\begin{equation}
  		x_{d}(t)=c\,{{\rm cn}(ct,k)\over{{\rm sn}(ct,k)}},\quad 
  		y_{d}(t)=-c^{2}{{\rm dn}(ct,k)\over{{\rm sn}^2(ct,k)}},
  			\quad z_{d}(t)=-\wp(t),%
  	\label{eq:functionsell4}
\end{equation}
in which $k^{\prime}=\sqrt{1-k^{2}}$.\\
A third choice is
\begin{eqnarray}
   		x(t)&=&{c\over2}\left[ {{{\rm dn}^{2}(ct/2,k)+ikk^{\prime}
   {\rm sn}^2(ct/2,k)}
   		\over {\rm sn}(ct/2,k)\,{\rm dn}(ct/2,k)}+
   {k\,{\rm cn}^{2}(ct/2,k)-ik^{\prime}
   \over{{\rm cn}(ct/2,k)}}\right],\nonumber\\
   y(t)&=&x^\prime(t),
   			\quad z(t)=-\wp(t),
   	\label{eq:functionsell5}
\end{eqnarray}
and
\begin{equation}
   		x_{d}(t)=c\,{{\rm dn}(ct,k)\over{{\rm sn}(ct,k)}},\quad 
   		y_{d}(t)=-c^{2}{{\rm cn}(ct,k)\over{{\rm sn}^2(ct,k)}},
   			\quad z_{d}(t)=-\wp(t).%
   	\label{eq:functionsell6}
\end{equation}
For the elliptic Lax pair with spectral parameter we find several
sets of functions which are closely related to each other.
The first set is:
\begin{eqnarray}
   x(t,\xi)&=&{\sigma({\xi/2}-
   t)\over{\sigma({\xi/2})\sigma(t)}},
   \quad  y(t,\xi)= x(t,\xi)\left[\zeta(
   t-\xi/2)-\zeta(t)\right],\nonumber\\
   && \hspace{3cm} z(t,\xi)=-\left[\wp(
   t)-\wp({\xi/2})\right],\label{simpsolx}\\ 
   x_{d}(t,\xi) &=&
   {\sigma(\xi-t)\over{\sigma(\xi)\sigma(t)}},
   \quad \ \ y_d(t,\xi)= x_d(t,\xi)\left[\zeta(
   t-\xi)-\zeta(t)\right],\nonumber\\
   && \hspace{2.9cm} z_d(t,\xi)=-\left[\wp(t)-\wp(\xi)\right].
   \label{simplesol}
\end{eqnarray}
in which \(\sigma\) and \(\zeta\) are Weierstrass' sigma and zeta 
functions. The other sets of functions are related to the above one
by simple shifts of the parameter \(\xi\) and the `gauge
transformation'  (\ref{gaugefr}) explained below:
\begin{eqnarray}
    x(t,\xi)&=&{\sigma({\xi/2}+\omega_j-
    t)\over{\sigma({\xi/2}+\omega_j)\sigma(t)}}
    \exp[t(\eta_j+\zeta(\xi)/2)],\qquad \quad \ \
    \eta_j=\zeta(\omega_j),\nonumber\\
     y(t,\xi)&=&x^\prime(t,\xi),\quad z(t,\xi)=-[\wp(
    t)-\wp({\xi/2}+\omega_j)], \quad j=1,2,3,\label{simplesol2}\\
    x_{d}(t,\xi) &=& {\sigma(\xi-t)\over{\sigma(\xi)\sigma(
t)}}
    \exp[t\zeta(\xi)],\nonumber\\
     y_d(t,\xi)&=&x_d^\prime(t,\xi),\quad
     z_d(t,\xi)=-\left[\wp(t)-\wp(\xi)\right].
    \label{simplesol2xd}
\end{eqnarray}

In the trigonometric and hyperbolic functions the constant
\(a\) is a free parameter  setting the scale
of the theory.
One obtains the rational potential in the limit \(a\to0\).
 The trigonometric  ($k\to0$)  and hyperbolic ($k\to1$) limits of
the elliptic cases give other sets of functions for these cases. 
One important property is that they all satisfy the {\em sum rule}
\begin{equation}
   	y(u)x(v)-y(v)x(u)=
   	x(u+v)[z(u)-z(v)],
   	\quad u,v\in {\mathbb C}.
   	\label{eq:ident1}
\end{equation}
The functions $x_d$, $y_{d}$ and $z_d$ satisfy the same relations,
including those containing the spectral parameter.
These functions also satisfy  a {\em second sum rule}
\begin{eqnarray}
   x(-v)\,y(u)-x(u)\,y(-v)
   &+&2\left[x_d(u)\,y(-u-v)-
   y(u+v)\,x_d(-v)\right]\nonumber\\
   &+&x(u+v)\,y_d(-v)-y_d(u)\,x(-u-v)=0,
   	\label{eq:xydiden}
\end{eqnarray}
which  is  essentially the same as the condition (3.29)
of a previous paper \cite{bcs}.
In all of these cases the inter-particle potential \(V\)
is proportional to
\(-z + const\) (see the Hamiltonian
(\ref{eq:hamiltonian}))  and $y$ ($y_{d}$) is the
derivative of
$x$ ($x_d$) and 
$z$ is always an even function:
\begin{equation}
   	y(t)=x^\prime(t),\quad z(t)=x(t)x(-t)+constant,\quad
   	 z(-t)=z(t).
   	\label{eq:parity}
\end{equation}
It should be remarked that the set of functions \(\{x(t),
x_d(t)\}\) has a kind of `gauge freedom'. If \(\{x(t),
x_d(t)\}\) satisfies the first and the second sum rules,
then
\begin{equation}
   \{ \tilde{x}(t)=x(t)e^{tb},\quad
   \tilde{x}_d(t)=x_d(t)e^{2tb}\}
   \label{gaugefr}
\end{equation}
also satisfies the same sum rules. Here \(b\) is an arbitrary 
\(t\)-independent constant, which can depend on \(\xi\). The
function \(z(t)\) is gauge invariant.

  For the rational (\ref{eq:functions}), trigonometric 
(\ref{eq:functionstri}) and hyperbolic cases 
(\ref{eq:functionshyp})
$x$ is an odd function and $y$ is 
an  even function 
but they do not have definite parity for the elliptic potentials 
(\ref{eq:functionsell1}) -- (\ref{simplesol2}). 
The Hamiltonian (\ref{eq:hamiltonian}) is proportional to
the lowest  conserved quantity  up to a 
constant:
\begin{equation}
   	Tr(L^2)=2I_{Adj}{\cal H}=4h{\cal H},
   	\label{eq:consham}
\end{equation}
in which \(I_{Adj}\) is the
{\em second Dynkin index} for the adjoint representation and
$h$ is the Coxeter number.

\subsection{Minimal type Lax pair}
The minimal type  Lax pair  is represented in the set of
weights of a minimal representation 
\begin{equation}
   	\Lambda=\{\mu,\nu,\rho, \ldots \},
   	\label{eq:minweight}
\end{equation}
of a semi-simple {\em simply laced} algebra
\({\mathfrak g}\)
 with root system \(\Delta\)
of rank \(r\). 
It is characterised \cite{bcs,GoOl} by the condition that any
weight
\(\mu\in\Lambda\) has scalar products with the roots restricted
as follows: 
\begin{equation}
   	{2\alpha\cdot\mu\over{\alpha^2}}=0,\pm 1,
    \quad \quad \forall\mu\in\Lambda
   	\quad \ \mbox{and}\quad \forall\alpha\in\Delta.
   	\label{eq:mindef}
\end{equation}
Corresponding to (\ref{rootscheme}), we have the following
universal pattern for the minimal representations:
\begin{equation}
   \mbox{Minimal Representation}:\qquad \mbox{ weight}
   - \mbox{ weight}=\left\{
   \begin{array}{l}
      \mbox{root}\\
      \mbox{non-root}
   \end{array}
   \right.
   \label{minscheme}
\end{equation}
Due to the definition of the minimal weights (\ref{eq:mindef}) 
there can be no terms like \(2\times\) root etc. in the right
hand side of (\ref{minscheme}).
This determines the  structure of the minimal type Lax pairs:
\begin{eqnarray}
   	L(q,p,\xi) & = & p\cdot H + X ,\nonumber\\
   	M(q,\xi) & = & D+Y.
   	\label{eq:minLaxform}
\end{eqnarray}
The matrices \(H\), \(X\) and \(Y\) have the same form as before
\begin{equation}
   	X=ig\sum_{\alpha\in\Delta}x(\alpha\cdot
    q, \xi)E(\alpha),\quad
   	Y=ig\sum_{\alpha\in\Delta}y(\alpha\cdot q, \xi)E(\alpha),
   	\label{eq:minXYdef}
\end{equation}
corresponding to the first line of (\ref{minscheme}).
We need only functions $x$, $y$ and $z$ (no $x_{d}$ etc.) and
they  need only satisfy (\ref{eq:ident1}) but not
(\ref{eq:xydiden}). Thus, besides those listed in section two 
(\ref{eq:functions})-(\ref{eq:functionsell6}), 
there are more choices of these functions, for example 
\cite{OP1}:
\begin{equation}
   	x(t)={a\over{\sin at}},\quad {a\over{\sinh at}}, 
   	\quad a: const.\quad {c\over{{\rm sn}(ct,k)}},
   	\quad c{{\rm cn}(ct,k)\over{{\rm sn}(ct,k)}},
   	\quad c{{\rm dn}(ct,k)\over{{\rm sn}(ct,k)}},
    \quad c=\sqrt{e_1-e_3}
   	\label{eq:morepoten}
\end{equation}
for the trigonometric, hyperbolic and elliptic potentials.
As in the case of the root type Lax pair, the three choices
of functions for the elliptic potentials are related with
each other by modular transformations.
 For the elliptic Lax
pair with spectral parameter 
\cite{Kri,DHPh}:
\begin{eqnarray}
   x(t,\xi) &=&{\sigma(\xi- t)\over
   {\sigma(\xi)\sigma(t)}},\qquad
   y(t,\xi)=
x(t,\xi)\left[\zeta(t-\xi)-\zeta(t)\right],
   \nonumber\\
   z(t,\xi)&=&-\left(\wp(t)-\wp(\xi)\right).
   \label{eq:specfun3}
\end{eqnarray}
The difference with the root type Lax pair 
is that their matrix elements are labeled by the 
 weights instead of the roots:
\[
    H_{\mu \nu}=\mu\delta_{\mu, \nu},\quad 
   	E(\alpha)_{\mu \nu}=\delta_{\mu-\nu,\alpha}.
\]
In the diagonal matrix \(D\) the terms related to 
the double roots are
dropped:
\begin{equation}
   	D_{\mu \nu}= \delta_{\mu, \nu}D_{\mu},\quad
   	D_{\mu}=-ig\sum_{\Delta\ni\beta=\mu-\nu}z(\beta\cdot q).
   	\label{eq:minD}
\end{equation}
Here the summation is over roots \(\beta\) such that for 
\(\exists\nu\in\Lambda\)
\[
   \mu-\nu=\beta\in\Delta.
\]
The Hamiltonian 
(\ref{eq:hamiltonian}) is  proportional to the lowest
conserved quantity  for the minimal type Lax pair, too:
\begin{equation}
   	Tr(L^2)=2I_{\Lambda}{\cal H}.
   	\label{eq:minconsham}
\end{equation}
Here $I_{\Lambda}$ is  the second Dynkin index
(\ref{eq:consham}) of the representation \(\Lambda\).
For the proof of the equivalence of the Lax equation 
\[
   \dot{L}={d\over{dt}}L=[L,M]
\]
\bigskip
and the canonical equation for the Hamiltonian
(\ref{eq:hamiltonian}) see our previous paper
\cite{bcs}.

\section{Symmetries and  Reductions of  Elliptic
Calogero-Moser Models}
\setcounter{equation}{0}

In this section we discuss  the symmetries of the
Calogero-Moser models  with the elliptic potential:
\begin{equation}
   	{\cal H}={1\over2}p^2+{g^2\over2}\sum_{\alpha\in\Delta}
   	\wp(\alpha\cdot q)
   	\label{eq:hamiltonianell}
\end{equation}
based on the root system \(\Delta\) of a semi-simple simply
laced algebra \(\mathfrak{g}\).
As is well known, the root system \(\Delta\) is
characterised by its Dynkin diagram.
The Dynkin diagrams (and the extended ones with the affine
root attached) of simply laced algebras have various {\em
automorphisms} \(A\), which map a root to another:
\begin{equation}
   A\alpha\in\Delta,\quad \forall\alpha\in\Delta.
   \label{auto}
\end{equation}
Thus by combining transformation by an automorphism with the
periodicity of the elliptic potential,
we find that the above Hamiltonian (\ref{eq:hamiltonianell})
is invariant under the following discrete transformation of the
dynamical variables:
\begin{eqnarray}
   q&\to& q^\prime=Aq+{2\omega}\lambda,\nonumber\\
   p&\to& p^\prime=Ap,
   \label{autoshift}
\end{eqnarray}
in which \(2\omega\) is any one of the periods
(\(2\omega_1,2\omega_2,2\omega_3\)) of the Weierstrass elliptic
function \(\wp\) and \(\lambda\) is an arbitrary element of the
{\em weight lattice} 
. That is, it satisfies
\[
   \alpha\cdot\lambda\in{\mathbb Z},\quad \forall\alpha\in\Delta.
\]
By restricting the dynamical variables to the {\em invariant
subspace} of the transformation
\begin{eqnarray}
   q&=&Aq+{2\omega}\lambda,\nonumber\\
   p&=&Ap,
   \label{invspeq}
\end{eqnarray}
we obtain a {\em reduced model} of the elliptic Calogero-Moser
model. In terms of the roots this corresponds to folding
the simply laced root system \(\Delta\) by the automorphism 
\(A\) (see \cite{folding} for the corresponding examples in Toda
theories). If we choose the automorphism of the ordinary
(un-extended) Dynkin diagram
\(A_u\), an untwisted non-simply laced root system is obtained. For
the automorphism of the extended Dynkin diagram \(A_e\) one obtains
a twisted non-simply
laced root system. 
According to the nature of the automorphism
\(A\) and the choice of \(\lambda\) we have the following
two different cases:\\
\smallskip
(i) \textbf{Untwisted non-simply laced model}. We choose the
automorphism of the un-extended Dynkin diagram \(A_u\) and
\(\lambda\equiv0\). Since
\(\lambda\equiv0\), the periodicity is irrelevant and the model is
defined for all four choices of the potential. This gives the well
known Calogero-Moser models based on
untwisted non-simply laced root systems. 
Various examples of this reduction for minimal type
Lax pairs were presented in a previous paper \cite{bcs}.
The Lax pair for these reduced models with as many independent coupling
constants as  independent Weyl orbits in the root system will
be fully discussed in section 4. \\
\smallskip
(ii) \textbf{Twisted non-simply laced model.} Let us choose the 
automorphism of the extended Dynkin diagram \(A_e\) and
some special weight  \(\lambda\) 
(in most cases it is a
minimal weight \(\lambda_{min}\) or a linear combination of
them). In this case some of the roots vanish in the
invariant subspace
\cite{folding,bcds}. In order to avoid the singularity of the
elliptic function, a non-vanishing weight vector \(\lambda\)
is necessary and it should have non-vanishing scalar products with
the roots which are mapped to zero. Some of these models have been
introduced in
\cite{DHPh} in a different context. 
A twisted \(BC_n\)  model (\ref{twa2nham}) is obtained by the folding
\(D_{2n+2}^{(1)}\to A_{2n}^{(2)}\). 
(In this paper we use notation like \(D_{2n}^{(1)}\) only to
indicate the extended Dynkin diagram but not the affine Lie algebra.)
This will be derived in the
subsection 3.5. One might be tempted to combine
the automorphism of unextended Dynkin diagram
\(A_u\) with a non-vanishing weight \(\lambda\).
So far as we have tried this does not lead to a new 
integrable model.

\bigskip
It should be noted that all of the models derived in this
section are a subsystem of the Calogero-Moser models based on
simply laced root systems. Thus the integrability of these
models is inherited from the original models. 

\bigskip
In the rest of this section we consider the reduction by
automorphisms
\(A_e\) of the extended Dynkin diagrams. The corresponding
reductions
 of
the Dynkin diagrams are:
\begin{equation}
    D_{2n}^{(1)}\to A_{2n-1}^{(2)}, \quad D_{n+2}^{(1)}\to
   D_{n+1}^{(2)},\quad E_7^{(1)}\to E_6^{(2)}, 
   \quad  E_6^{(1)}\to
   D_4^{(3)}  
    \quad \mbox{and}
   \quad D_{2n+2}^{(1)}\to A_{2n}^{(2)}.
   \label{twistfold}
\end{equation}
These automorphisms satisfy (we denote by \(A\) for brevity)
\begin{equation}
   A^2=1\quad \mbox{for} \quad  D_{2n}^{(1)}\to
   A_{2n-1}^{(2)},\quad D_{n+2}^{(1)}\to
   D_{n+1}^{(2)},
   \quad E_7^{(1)}\to E_6^{(2)},
   \label{autodeg3}
\end{equation} 
\begin{equation}
   A^3=1\quad \mbox{for}  \quad E_6^{(1)}\to
   D_4^{(3)}, \qquad \quad A^4=1 \quad \mbox{for} \quad
   D_{2n+2}^{(1)}\to A_{2n}^{(2)}.
   \label{autodeg4}
\end{equation}
For these automorphisms we consider the equation (\ref{invspeq})
determining the invariant subspace of the discrete transformation
(\ref{autoshift}). The projector to the invariant subspace is
given by
\begin{equation}
   Pr={1\over2}(1+A),\qquad Pr={1\over3}(1+A+A^2) \quad \mbox{and}
   \quad Pr={1\over4}(1+A+A^2+A^3), 
   \label{proj2}
\end{equation}
respectively.
By multiplying the first 
equation  determining
the invariant subspace (\ref{invspeq}) by  \(A\) (and
\(A^2\)),  we obtain
\[
   Aq=A^2q+{2\omega}A\lambda,
   \quad A^2q=A^3q+{2\omega}A^2\lambda.
\]
This puts a restriction on the possible choice of the
weight vector \(\lambda\):
\begin{equation}
   Pr\lambda=0.
   \label{lamres}
\end{equation}
Let us consider the reductions listed in (\ref{twistfold}) in
turn.

\subsection{Twisted \(C_n\) model}
The Dynkin diagram of  \(A_{2n-1}^{(2)}\) is
obtained from that of \(D_{2n}^{(1)}\) by the following
folding:\\

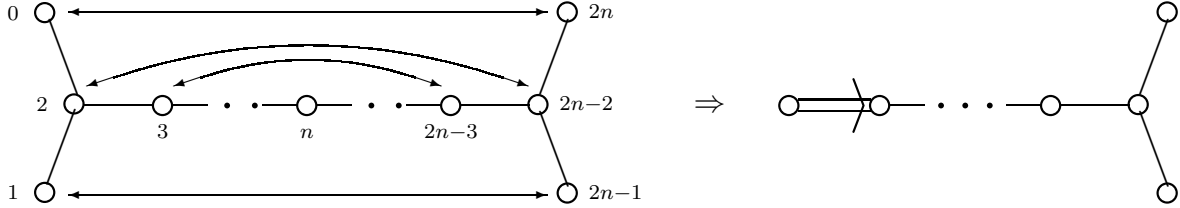
\begin{picture}(400,80)
   \put(5,40){$\edddnds{0}.{2}.{1}.{3}.{n}.
   {2n-3}.{2n-2}.
   {2n-1}.{2n}.~\Rightarrow~ 
   \eddanod{}.{}.{}.{}.{}.{}.$}
   \put(73,53){\vector(-3,-1){10}}
   \put(154,53){\vector(3,-1){10}}
   \qbezier(73,53)(113.5,66.5)(154,53)
   \put(40,53){\vector(-3,-1){10}}
   \put(187,53){\vector(3,-1){10}}
   \qbezier(40,53)(113.5,77.5)(187,53)
   \put(197,78){\vector(1,0){7}}
   \put(30,78){\vector(-1,0){7}}
   \put(30,78){\line(1,0){167}}
   \put(30,8.5){\vector(-1,0){7}}
   \put(197,8.5){\vector(1,0){7}}
   \put(30,8.5){\line(1,0){167}}
\end{picture}

The \(C_n\) Dynkin diagram is contained in the
\(A_{2n-1}^{(2)}\) Dynkin diagram.
The automorphism is given by
\begin{equation}
    A\alpha_j=\alpha_{2n-j},\quad j=0,1,\ldots,2n,
\end{equation}
in which \(\{\alpha_j\}\), \(j=1,\ldots,2n\) are \(D_{2n}\) 
simple
roots in an orthonormal basis of \({\mathbb R}^{2n}\):
\begin{equation}
    \alpha_1=e_1-e_2,\quad \cdots,\ 
    \alpha_{2n-2}=e_{2n-2}-e_{2n-1},\quad
    \alpha_{2n-1}=e_{2n-1}-e_{2n} 
    \quad \alpha_{2n}=e_{2n-1}+e_{2n}
    \label{dnsimroot}
\end{equation}
and \(\alpha_0=-(e_1+e_2)\).
In terms of the orthonormal basis the automorphism \(A\) has
a simple expression:
\begin{equation}
   Ae_j=-e_{2n+1-j},\quad j=1,\ldots,2n.
\end{equation}
Among the \(2n(4n-2)\) roots of \(D_{2n}\), \(2n\) roots
\begin{equation}
   \pm(e_j-e_{2n+1-j}),\quad j=1,\ldots,n
\end{equation}
belong to the invariant subspace of \(A\). That is these
\(2n\) roots remain long roots after folding. There are
\(2n\) roots which are eigenvectors of \(A\) with eigenvalue
-1:
\begin{equation}
   \pm(e_j+e_{2n+1-j}),\quad j=1,\ldots,n,
   \label{cnzeroroots}
\end{equation}
\begin{equation}
   A(e_j+e_{2n+1-j})=-(e_j+e_{2n+1-j}).
\end{equation}
These  \(2n\) roots are mapped to the null vector in the
invariant subspace. The remaining \(8n(n-1)\) roots are
mapped to the \(2n(n-1)\) short roots of \(C_n\) four to one.
In fact (\(j,k=1,\ldots,n\))
\begin{equation}
   \begin{array}{lcclc}
      (a)& e_j+e_k &\hspace{1cm} &(b)& -e_{2n+1-j}+e_k\\
      (c)& -e_{2n+1-k}+e_j & \hspace{1cm}&(d)
      &-e_{2n+1-j}-e_{2n+1-k}
   \end{array}
   \label{cnfour}
\end{equation}
are all mapped to a short root
\begin{equation}
   Pr(e_j+e_k)={1\over2}(e_j-e_{2n+1-j}+e_k-e_{2n+1-k}),
\end{equation}
which has (length)$^2=1$. There is a unique minimal weight
(the spinor weight \(\lambda_{2n}\)) 
which is annihilated by \(Pr\):
\begin{equation}
   Pr\lambda_{2n}={1\over2}Pr(e_1+e_2+\cdots+e_{2n})=0.
\end{equation}
As expected \(\lambda_{2n}\) has a scalar product 1 (mod 2) with
all the roots (\ref{cnzeroroots}) which are mapped to the
origin of the invariant subspace. Thus the singularity 
of the elliptic potential is avoided after folding.
It is easy to see that \(\lambda_{2n}\) has a scalar product 1
(mod 2) with two of the short roots in (\ref{cnfour}) and
scalar product 0 (mod 2) with the other two. 
It is easy to see that the solution of (\ref{invspeq}) is given by
\begin{eqnarray}
   q&=&\sum_{j=1}^nQ^jv_j+{\omega}\lambda_{2n},\quad
   v_j={1\over{\sqrt2}}(e_j-e_{2n+1-j}),
   \label{invspsoltwcn}\\
   p&=&\sum_{j=1}^n P_jv_j,\qquad \qquad \ 
   \lambda_{2n}=\sum_{j=1}^n{1\over2}(e_1+
   \cdots+e_{2n-1}+e_{2n}),
   \label{lambsol}
\end{eqnarray}
in which \(\{Q^j,P_j\}\) are the  canonical
variables for the reduced system.
By substituting the above solution into the original
Hamiltonian we arrive at the twisted \(C_n\)
Calogero-Moser model
\begin{equation}
   	{\cal
    H}={1\over2}\sum_{j=1}^nP_j^2+{g^2\over2}
    \sum_{\alpha\in\Delta_l}
   	\wp(\alpha\cdot Q)+{g^2}\sum_{\mu\in\Delta_s}
   	\left[\wp(\mu\cdot Q)+\wp(\mu\cdot Q+\omega)\right],
   	\label{eq:twcnellham}
\end{equation}
in which the sets of long and short roots are:
\begin{equation}
   \Delta_l=\{\pm \sqrt2 v_j:\ j=1,\ldots,n\},\qquad
   \Delta_s=\{{1\over{\sqrt2}}(\pm v_j\pm v_k):\ j,k=1,\ldots,n\}.
   \label{cnroots}
\end{equation}
It is well known that the combination of elliptic functions
appearing in the short root potential can be expressed in terms
of an elliptic function with a half  period.
 For example, the \(\wp^{(1/2)}\) function
\[
    \wp^{(1/2)}(x)\equiv \wp(x)+\wp(x+\omega_1)-\wp(\omega_1)
\]
has the set of fundamental periods \(\{\omega_1,2\omega_3\}\)
instead of the
original \(\{2\omega_1,2\omega_3\}\).

\subsection{Twisted \(B_n\) model}
The Dynkin diagram of  \(D_{n+1}^{(2)}\) is
obtained from that of \(D_{n+2}^{(1)}\) by the following
folding:\\
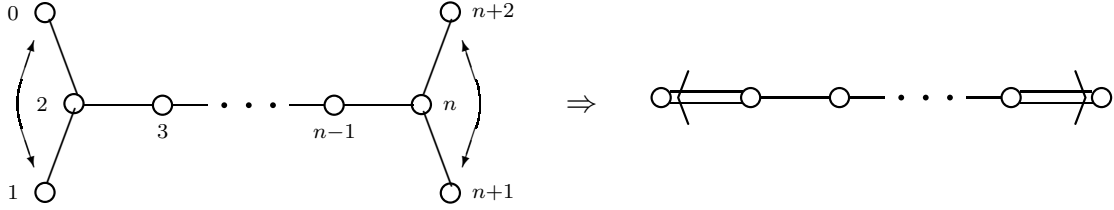
\begin{picture}(400,90)
   \put(20,40){$\edddnd{0}.{2}.{1}.{3}.{n-1}.
   {n}.{n+1}.{n+2}.~\Rightarrow~
   \edddniid{}.{}.{}.{}.{}.$}
   \put(192,52){\vector(-1,3){5}}
   \put(192,34){\vector(-1,-3){5}}
   \qbezier(192,52)(195,43)(192,34)
   \put(20,52){\vector(1,3){5}}
   \put(20,34){\vector(1,-3){5}}
   \qbezier(20,52)(17,43)(20,34)
\end{picture}

The \(B_n\) Dynkin diagram is contained in the
\(D_{n+1}^{(2)}\) Dynkin diagram.
The automorphism is given by
\begin{equation}
   \begin{array}{ccc}
      A\alpha_0=\alpha_{1},&A\alpha_{n+1}=\alpha_{n+2},
      &A\alpha_j=\alpha_{j},\quad  j=2,\ldots,n,\\
      A\alpha_1=\alpha_{0},&A\alpha_{n+2}=\alpha_{n+1},&
   \end{array}
   \label{dn2auto}
\end{equation}
in which \(\alpha_j\), \(j=1,\ldots,n+2\) are 
\(D_{n+2}\) simple
roots and \(\alpha_0\) is the affine root. By using the
expression of the simple roots in terms of 
an orthonormal basis of \({\mathbb R}^{n+2}\) (see
(\ref{dnsimroot})) the automorphism \(A\) is expressed as
\begin{equation}
   Ae_1=-e_{1},\quad Ae_{n+2}=-e_{n+2},\quad Ae_j=e_j,\quad 
   j=2,\ldots,n+1.
   \label{dn2auto2}
\end{equation}
This means that the invariant subspace 
of the automorphism \(A\)
is spanned by \(e_j\), (\(j=2,\ldots,n+1\)) and the two
dimensional subspace spanned by \(\{e_1,e_{n+2}\}\) is
annihilated by the projector \(Pr\):
\begin{eqnarray}
   Pr\,e_j&=&e_j,\quad j=2,\ldots,n+1,\label{bntwpr}\\
   Pr\,e_1&=&Pr\,e_{n+2}=0.
   \end{eqnarray}
Among the \(2n(n+2)(n+1)\) roots of \(D_{n+2}\), the following
\(2n(n-1)\) roots
\begin{equation}
   \pm e_j\pm e_k,\quad j,k=2,\ldots,n+1,
   \label{bntwlong}
\end{equation}
remain long and become the long roots of \(B_n\). 
There are four
roots which are mapped to the origin of the invariant subspace:
\begin{equation}
   \pm e_1\pm e_{n+2}.
   \label{bntwzero}
\end{equation}
The remaining \(8n\) roots are mapped to short roots four
to one:
\begin{equation}
   \left.
   \begin{array}{r}
      \pm e_1\pm e_j\\
      \pm e_{n+2}\pm e_j
   \end{array}\right\}\to \pm e_j,\quad j=2,\ldots,n+1.
\end{equation}
It is easy to see that
\begin{eqnarray}
   q&=&Q+{\omega}\lambda,\qquad Q=\sum_{j=2}^{n+1}q^je_j,
   \label{invspsolsbn}\\
   p&=&P,\qquad \qquad \ P=\sum_{j=2}^{n+1} p_je_j,\nonumber\\
   \lambda&=&\lambda_1=e_{1},\qquad
   \mbox{or}\quad \lambda= e_{n+2}=\lambda_{n+2}-\lambda_{n+1},
   \label{invspsolsbnlamb}
\end{eqnarray}
are solutions of (\ref{invspeq}).  In other words, 
this means that
\begin{equation}
   \{q^1=0,\quad q^{n+2}=\omega,\quad
   \mbox{or}\quad q^1=\omega,\quad q^{n+2}=0\}\quad \mbox{and}
   \quad p_{1}=0,\quad  p_{n+2}=0
   \label{bnrest2}
\end{equation}
are  valid restrictions of the elliptic \(D_{n+2}\) Calogero-Moser
model. The above \(\lambda\) (\ref{invspsolsbnlamb}) has a
non-vanishing scalar products with the roots which are mapped to
zero:
\[
   \lambda\cdot(\pm e_1\pm e_{n+2})=1\quad \mbox{mod} \quad  2.
\]
It has a scalar product 1 (mod 2) with one half of the roots 
which are mapped to short roots:
\[
   \lambda_1\cdot(\pm e_1\pm e_{j})=1\quad \mbox{mod} \quad  2
\]
and zero with the rest:
\[
   \lambda_1\cdot(\pm e_{n+2}\pm e_{j})=0.
\]
By substituting the solution (\ref{invspsolsbn}) into the
Hamiltonian, we obtain
\begin{equation}
   	{\cal
   H}={1\over2}\sum_{j=2}^{n+1}p_j^2+{g^2\over2}
   \sum_{\alpha\in\Delta_l}
   	\wp(\alpha\cdot q)+{g^2}\sum_{\mu\in\Delta_s}
   	\left[\wp(\mu\cdot q)+\wp(\mu\cdot
   q+\omega)\right]+const,
   	\label{eq:twbnellham}
\end{equation}
in which the sets of long and short roots are:
\begin{equation}
   \Delta_l=\{\pm e_j\pm e_k:\ j,k=2,\ldots,n+1\},\qquad
   \Delta_s=\{\pm e_j:\ j=2,\ldots,n+1\}.
   \label{bnroots2}
\end{equation}

\subsection{Twisted \(F_4\) model}
The Dynkin diagram of  \(E_{6}^{(2)}\) is
obtained from that of \(E_{7}^{(1)}\) by the following
folding:\\
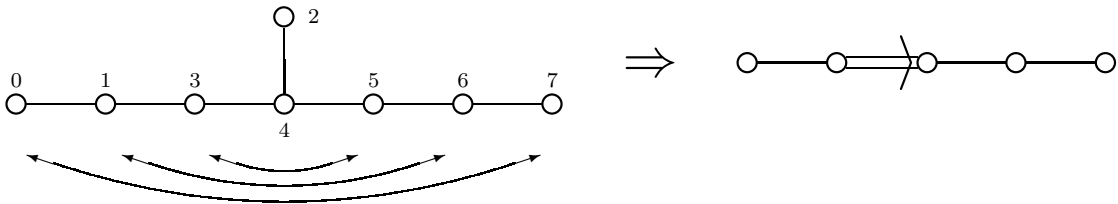
\begin{picture}(400,130)
   \put(20,90){$\eddeii{1}.{2}.{3}.{4}.{5}.
   {6}.{7}.{0}.~\Rightarrow~ 
   \eddfiiu{}.{}.{}.{}.{}.$}
   \put(225,57){\vector(3,1){10}}
   \put(51,57){\vector(-3,1){10}}
   \qbezier(51,57)(138,28)(225,57)
   \put(189,57){\vector(3,1){10}}
   \put(87,57){\vector(-3,1){10}}
   \qbezier(189,57)(138,40)(87,57)
   \put(156,57){\vector(3,1){10}}
   \put(120,57){\vector(-3,1){10}}
   \qbezier(120,57)(138,51)(156,57)
\end{picture}

The \(F_4\) Dynkin diagram is  contained in the \(E_6^{(2)}\)
Dynkin diagram.
As indicated in the diagram, the automorphism \(A\) is
given by:
\begin{equation}
   \begin{array}{lll}
      A\alpha_1=\alpha_6,\qquad
      &A\alpha_2=\alpha_2,\qquad& A\alpha_3=\alpha_5,\\
      A\alpha_4=\alpha_4,&A\alpha_5=\alpha_3,
      &A\alpha_6=\alpha_1,\\
      A\alpha_7=\alpha_0,&A\alpha_0=\alpha_7.&
   \end{array}
   \label{f4auto}
\end{equation}
Let us adopt the following representation of the 
simple roots of \(E_7\) in
terms of an orthonormal basis of \({\mathbb R}^7\):
\begin{equation}
   \begin{array}{ll}
      \alpha_1={1\over2}(e_1-e_2-e_3-e_4-e_5-e_6+\sqrt2
      e_7),&\alpha_2=e_1+e_2,\\ 
      \alpha_3=-e_1+e_2,&\alpha_4=-e_2+e_3,\\
      \alpha_5=-e_3+e_4,&\alpha_6=-e_4+e_5,\\
      \alpha_7=-e_5+e_6,&\alpha_0=-\sqrt{2}e_7.
   \end{array}
   \label{e7sipmroot}
\end{equation}
By a similar analysis as before, 
we find that among the 126 roots of \(E_7\)
the following 24 roots remain long:
\begin{equation}
   \begin{array}{llll}
      \pm(e_1+e_2),&\pm(e_2-e_3),&\pm(e_3+e_4),&
      {1\over2}(\pm(e_1+e_2)\pm(e_3+e_4)\pm(e_5-e_6+\sqrt2
      e_7)),\\
      \pm(e_1+e_3),&\pm(e_2+e_4),&\pm(e_1-e_4),
      &
      {1\over2}(\pm(e_1-e_4)\mp(e_2-e_3)\pm(e_5-e_6+\sqrt2 e_7)).
   \end{array}
   \label{f4long}
\end{equation}
The following 6 roots are mapped to 0:
\begin{equation}
   \pm(e_5+e_6),\quad
   {1\over2}(\pm(e_1-e_2-e_3+e_4)\pm(e_5-e_6+\sqrt2 e_7)).
   \label{f4zero}
\end{equation}
The remaining 96 roots are mapped to \(F_4\) 
short roots
four to one. It is easy to see that the solution of
(\ref{invspeq}) is given by
\begin{eqnarray}
    q&=&\sum_{j=1}^4Q^jv_j+{\omega}\lambda, \quad
    \lambda=\lambda_7=e_6+{1\over{\sqrt2}}e_7,
    \label{invspsoltwf4}\\
    p&=&\sum_{j=1}^4 P_jv_j, \qquad \ 
    \quad \mbox{or} \quad
    \lambda=\lambda_3-\lambda_5={1\over2}
    (-e_1+e_2+e_3-e_4-e_5-e_6),
    \label{f4lambsol}
\end{eqnarray}
in which \(\{v_j\}\), \(j=1,\ldots,4\) is a new orthonormal
basis of the four-dimensional invariant subspace:
\[
   \begin{array}{ll}
      v_1={1\over{\sqrt2}}(e_1+e_2),&
      v_2={1\over{\sqrt6}}(e_1-e_2+2e_3),
      \\
      v_3=
      {1\over{\sqrt{12}}}(-e_1+e_2+e_3+3e_4),&
      v_4={1\over2}(e_5-e_6+\sqrt2 e_7).
   \end{array}
\]
Both choices of \(\lambda\) have a scalar product 1 (mod 2) with 
all the roots (\ref{f4zero}) which are mapped to zero.
It is straightforward to check that both choices of \(\lambda\)
have a scalar product 1 (mod 2) with one half of the short
roots and 0 with the rest. By substituting the above solution
into the original Hamiltonian we arrive at the twisted
\(F_4\) Calogero-Moser model
\begin{equation}
   	{\cal
   H}={1\over2}\sum_{j=1}^4P_j^2+{g^2\over2}
   \sum_{\alpha\in\Delta_l}
   	\wp(\alpha\cdot Q)+{g^2}\sum_{\mu\in\Delta_s}
   	\left[\wp(\mu\cdot Q)+\wp(\mu\cdot
   Q+\omega)\right]+const.
   	\label{eq:twf4ellham}
\end{equation}

\subsection{Twisted \(G_2\) model}
The Dynkin diagram of  \(D_{4}^{(3)}\) is
obtained from that of \(E_{6}^{(1)}\) by the triple
folding:\\

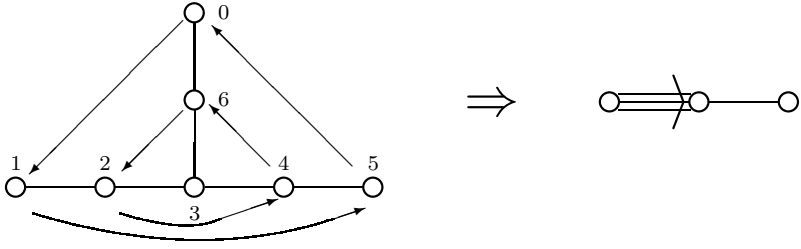
\begin{picture}(400,140)
\put(20,80){$\eddei{1}.{6}.{2}.{3}.{4}.
   {5}.{0}.\quad\Rightarrow\quad 
   \eddgiid{}.{}.{}.$}
   \put(114,41){\vector(3,1){22}}
   \qbezier(76,43)(102,35)(114,41)
   \put(157,41){\vector(3,1){12}}
   \qbezier(43,43)(106,24)(157,41)
   \put(133,61){\vector(-1,1){23}}
   \put(100,82){\vector(-1,-1){23}}
   \put(164,61){\vector(-1,1){53}}
   \put(100,116){\vector(-1,-1){58}}
\end{picture}\\
The \(G_2\) Dynkin diagram is  contained
in the \(D_4^{(3)}\) Dynkin diagram.
As indicated in the diagram, the automorphism \(A\) is
given by:
\begin{equation}
   \begin{array}{lll}
      A\alpha_1=\alpha_5,\qquad
      &A\alpha_2=\alpha_4,\qquad
      &A\alpha_3=\alpha_3,\\
      A\alpha_4=\alpha_6,&A\alpha_5=\alpha_0,
      &A\alpha_6=\alpha_2,\\
      A\alpha_0=\alpha_1.&&
   \end{array}
   \label{g2auto}
\end{equation}
Let us adopt the following representation of the 
simple roots of \(E_6\) in
terms of an orthonormal basis of \({\mathbb R}^6\):
\begin{equation}
   \begin{array}{ll}
      \alpha_1={1\over2}(e_1-e_2-e_3-e_4+e_5-\sqrt3
      e_6),\quad&\alpha_2=e_4-e_5,\\ 
      \alpha_3=e_3-e_4,&\alpha_4=e_4+e_5,\\
      \alpha_5={1\over2}(e_1-e_2-e_3-e_4-e_5+\sqrt3
      e_6),&\alpha_6=e_2-e_3,\\
      \alpha_0=-(e_1+e_2).&
   \end{array}
   \label{e6sipmroot}
\end{equation}
By a similar analysis as before, we find that among the 72
roots of
\(E_6\) the following 6 roots remain long:
\begin{equation}
   \pm(e_2+e_3),\quad \pm(e_2+e_4),\quad \pm(e_3-e_4).
   \label{g2long}
\end{equation}
The following 12 roots are mapped to 0:
\begin{equation}
   \pm e_1\pm e_5,\quad
   {1\over2}(\pm e_1\pm(e_2-e_3-e_4)\pm e_5\pm\sqrt3 e_6).
   \label{g2zero}
\end{equation}
The remaining 54 roots are mapped to the 6 short roots of
\(G_2\) nine to one.
The short roots have (length)\({}^2\)=2/3 
because of the third order folding. 
The invariant subspace of the automorphism \(A\) is
spanned by two vectors:
\begin{equation}
   v_1={1\over{\sqrt2}}(e_3-e_4),\quad
   v_2={1\over{\sqrt6}}(2e_2+e_3+e_4).
   \label{g2bas}
\end{equation}
Let us consider the equation (\ref{invspeq}) determining
the invariant subspace with \(\lambda\) one of the minimal
weights \(\lambda_1\) (or \(\lambda_5\)):
\begin{eqnarray*}
   q&=&Aq+{2\omega}\lambda_1, \quad
   \lambda_1=e_1-{1\over{\sqrt3}}e_6,\quad
   \lambda_5=e_1+{1\over{\sqrt3}}e_6 
   \\ p&=&Ap.
\end{eqnarray*}
It is elementary to check that the following satisfies 
the above equation:
\begin{eqnarray}
   q&=&\sum_{j=1}^2Q^jv_j+{2\omega\over{3}}
   (\lambda_1+\lambda_5),\quad
   \lambda_1+\lambda_5=2e_1,
   \label{g2inveqsol}\\
   p&=&\sum_{j=1}^2P_jv_j,\nonumber
\end{eqnarray}
in which \(\{Q^j,P_j\}\) are the canonical variables of
the reduced system and
\(\{v_1,v_2\}\) are given in (\ref{g2bas}).
It should be noted that \(\lambda_1+\lambda_5=2e_1\) has 
a scalar product 1 and 2 (mod 3) with all the roots
(\ref{g2zero}) which are mapped to 0.
Moreover, it has a scalar product 0, 1 and 2 (mod 3) with
each one third of the roots which are mapped to \(G_2\)
short roots.
By substituting the solution (\ref{g2inveqsol}) to the
elliptic
\(E_6\) Calogero-Moser Hamiltonian, we obtain the twisted
\(G_2\) Hamiltonian:
\begin{eqnarray}
   	{\cal
   H}&=&{1\over2}\sum_{j=1}^2P_j^2 + {g^2\over2}
   \sum_{\alpha\in\Delta_l}
   	\wp(\alpha\cdot
   Q)
   \label{twg2ham}\\
   &&\hspace{1cm}+ {3g^2\over2}\sum_{\mu\in\Delta_s}
   	\left[\wp(\mu\cdot Q)+\wp(\mu\cdot
   Q+{2\omega\over3})+\wp(\mu\cdot
   Q+{4\omega\over3})\right]+const.\nonumber
   	\label{eq:twg2ellham}
\end{eqnarray}

\subsection{\(A_{2n}^{(2)}\) model or twisted \(BC_n\)
model}
This  model is associated with the twisted
affine algebra \(A_{2n}^{(2)}\). It is obtained by
folding the \(D_{2n+2}^{(1)}\) diagram using the fourth order
automorphism.
After rescaling the \(A_{2n}^{(2)}\) algebra has
\(2n\) long  and short roots of the form \(\{\pm
2e_j\}\),
\(\{\pm e_j\}\), \(j=1,\ldots,n\) and \(2n(n-1)\) middle
roots of the form \(\{\pm e_j\pm e_k\}\),
\(j,k=1,\ldots,n\). So it could be understood as 
a twisted \(BC_n\) model. This will provide a Lie
algebraic interpretation of the \(BC_n\) model, as we
will show presently. The Dynkin diagram of 
\(A_{2n}^{(2)}\) is obtained from that of
\(D_{2n+2}^{(1)}\) by the fourth order folding:\\
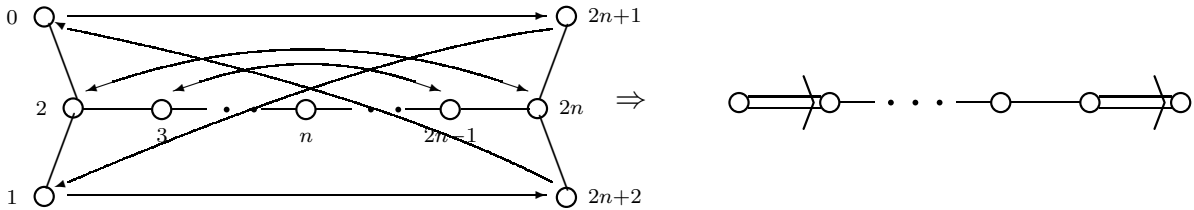
\begin{picture}(400,120)
   \put(5,60){$\edddnds{0}.{2}.{1}.{3}.{n}.
   {2n-1}.{2n}.{2n+2}.{2n+1}.~~ 
   \eddanidr{}.{}.{}.{}.{}.$}
   \put(73,73){\vector(-3,-1){10}}
   \put(154,73){\vector(3,-1){10}}
   \qbezier(73,73)(113.5,86.5)(154,73)
   \put(40,73){\vector(-3,-1){10}}
   \put(187,73){\vector(3,-1){10}}
   \qbezier(40,73)(113.5,97.5)(187,73)
   \put(197,98){\vector(1,0){7}}
   \put(23,98){\line(1,0){174}}
   \put(197,30){\vector(1,0){7}}
   \put(23,30){\line(1,0){174}}
   \put(23,35){\vector(-3,-1){5}}
   \qbezier(23,35)(140,85)(206,93)
   \put(23,93){\vector(-2,1){5}}
   \qbezier(23,93)(140,69)(206,35)
   \put(230,63){$\Rightarrow$}
\end{picture}

As shown above the \(D_{2n+2}\) root system is
invariant under the following automorphism:
\begin{equation}
   \begin{array}{ll}
      A\alpha_0=\alpha_{2n+1},\qquad&
      A\alpha_{1}=\alpha_{2n+2},\\
      A\alpha_{2n+1}=\alpha_{1},\qquad&
      A\alpha_{2n+2}=\alpha_{0},\\
      A\alpha_j=\alpha_{2n+2-j},\qquad&
      j=2,\ldots,2n.
   \end{array}
   \label{a2nauto}
\end{equation}
In terms of the standard orthonormal basis of
\({\mathbb R}^{2n+2}\) it is simply expressed as
\begin{equation}
   \begin{array}{ll}
      Ae_1=e_{2n+2},\qquad&
      Ae_{2n+2}=-e_{1},\\
      Ae_j=-e_{2n+3-j},\qquad&
      j=2,\ldots,2n+1.
   \end{array}
   \label{a2nauto2}
\end{equation}
That is, the automorphism \(A\) satisfies
\begin{equation}
   A^4=1,
   \label{a4}
\end{equation}
in the two-dimensional subspace spanned by
\(\{e_1,e_{2n+2}\}\) and in the rest of the space it 
satisfies
\begin{equation}
   A^2=1.
   \label{a42}
\end{equation}
Among the \(4(n+1)(2n+1)\) roots of
\(D_{2n+2}\) the following \(2n\) roots remain long:
\begin{equation}
   \pm(e_j-e_{2n+3-j}),\quad j=2,\ldots,n+1.
   \label{a2nlong}
\end{equation}
The following \(8n(n-1)\) roots are mapped to 
middle roots with (length)\({}^2\)=1:
\begin{equation}
   \pm e_j\pm e_k,\quad j+k\neq 2n+3,\quad 
   j,k=2,\ldots,2n+1.
   \label{a2nmid}
\end{equation}
In this case four different roots are mapped into one
middle root. There are \(16n\) roots that are mapped
to  short roots with (length)\({}^2\)=1/2:
\begin{equation}
   \pm e_1\pm e_j,\quad \pm e_{2n+2}\pm e_j\quad  
   j,k=2,\ldots,2n+1.
   \label{a2nshrt}
\end{equation}
In this case eight different roots are mapped into one
short root. Finally, there are \(2n+4\) roots which
are mapped to zero:
\begin{equation}
   \pm (e_j+e_{2n+3-j}),\quad \pm e_1\pm e_{2n+2}\quad  
   j,k=2,\ldots,n+1.
   \label{a2nzero}
\end{equation}
We look for  a solution of equation (\ref{invspeq}) 
 with \(\lambda=\lambda_{2n+1}\), the anti-spinor weight which
is a minimal weight:
\begin{eqnarray*}
   q&=&Aq+{2\omega}\lambda_{2n+1}, \quad
   \lambda_{2n+1}={1\over2}
   (e_1+\cdots+e_{2n+1}-e_{2n+2}),\\
   p&=&Ap.
\end{eqnarray*}
It is elementary to verify that 
\begin{eqnarray}
   q&=&\sum_{j=2}^{n+1}Q^jv_j+{\omega}\tilde\lambda,\quad 
   \tilde\lambda=e_1+{1\over2}(e_2+\cdots+e_{2n+1}),
   \label{a4nsol}\\
   p&=&\sum_{j=2}^{n+1}P_jv_j,
\end{eqnarray}
is a solution.
In the above expression, \(\{Q^j,P_j\}\), 
(\(j=2,\ldots,n+1\)) 
are the canonical variables of the reduced system and
\begin{equation}
   v_j={1\over{\sqrt2}}(e_j-e_{2n+3-j}),\quad j=2,\ldots,n+1,
\end{equation}
is an orthonormal basis of the invariant subspace of \(A\).
It is easy to see that \(\tilde\lambda\) has a vanishing scalar
product with all the long roots (\ref{a2nlong}). As with the
middle roots \(\tilde\lambda\) has a scalar product 1 (mod 2)
with one half of them and 0 with the rest.
An interesting situation arises when we consider the scalar
products of \(\tilde\lambda\) with the short roots
(\ref{a2nshrt}):
\begin{equation}
   \alpha\cdot\tilde\lambda=\left\{
   \begin{array}{llllclc}
      {1\over2}&\mbox{mod}&2&\mbox{for}&\alpha=\pm
      e_1-e_j&\mbox{and}&\pm e_{2n+2}+e_j,\\
      &&&&&&\\
      {3\over2}&\mbox{mod}&2&\mbox{for}&\alpha=\pm
      e_1+e_j&\mbox{and}&\pm e_{2n+2}-e_j.
   \end{array}
   \right.
   \label{a2nquart}
\end{equation}
It should be noted that \(\alpha\cdot\tilde\lambda=1,0\)
mod 2 do not occur for short roots \(\alpha\).
Finally \(\tilde\lambda\) has a scalar product 1 (mod 2) with
all the roots (\ref{a2nzero}) that are mapped to zero.

By substituting the above solution (\ref{a4nsol}) into the
Hamiltonian of elliptic \(D_{2n+2}\) Calogero-Moser model,
we obtain:
\begin{eqnarray}
   	{\cal
   H}&=&{1\over2}\sum_{j=2}^{n+1}P_j^2+{g^2\over2}
   \sum_{\Xi\in\Delta_l}
   	\wp(\Xi\cdot
   Q)
   \label{twa2nham}\\
   &&\hspace{1cm}+\ {g^2}\sum_{\alpha\in\Delta_m}
   	\left[\wp(\alpha\cdot Q)+\wp(\alpha\cdot
   Q+\omega)\right]\nonumber\\
   &&\hspace{1cm}+\ 2g^2\sum_{\mu\in\Delta_s}\left[\wp(
   \mu\cdot Q+ {\omega\over2})+\wp(\mu\cdot
   Q+{3\omega\over2})\right]+const,
   	\nonumber
\end{eqnarray}
in which \(\Delta_l, \Delta_m\) and \(\Delta_s\) are the sets of
long, middle and short roots of \(BC_n\) system, respectively.
This model was previously described in \cite{Ino}.

Before closing this subsection, some  remarks are
in order. First, the twisted models
derived in this subsection inherit the integrability of the
original simply laced models. The conserved quantities of
the twisted models are obtained from those of the simply
laced theory by  substitution of the variables.
Second, the Hamiltonian of the ordinary \(BC_n\) system
could be obtained from the Hamiltonian of \(D_{2n+2}\)
theory by the same folding as above with \(\lambda=0\), if
we ignore the singularities of the potential caused by the
vanishing roots. Third, as we have remarked at the 
beginning of this section, we have utilised only the automorphism
of the extended Dynkin diagram as relevant to the ordinary root
vectors.
The actual connection with the underlying affine algebras
is rather subtle, established only in the limit to the affine Toda
theories \cite{Ino,DHPh}.
However, the very fact that the affine root system 
(without the null roots) plays a fundamental role here seems to 
suggest the existence of an infinite dimensional algebra
(perhaps a kind of toroidal algebra \cite{GN})
in the elliptic Calogero-Moser systems.
This algebra is supposed to play the same or similar role
as are played by  affine algebras in the affine Toda theories.

\section{Independent Coupling Constants}
\setcounter{equation}{0}
In the previous section we have shown that all of the
Calogero-Moser models based on non-simply laced root systems,
the untwisted as well as the twisted models are obtained by folding
(reduction) of the models based on simply laced root systems.
These non-simply laced models inherit integrability as well as
restrictions from the original simply laced theories.
In these cases the ratio of the coupling constants for the long and
short root potentials are fixed by the order of the automorphism
used for the folding.
In fact these models are integrable even when these coupling
constants are independent.

In this section we will give the root type Lax pairs of the
untwisted non-simply laced models with as many independent
coupling constants as  independent Weyl
orbits in the set of roots. The independence of
the coupling constants stems from the 
independence of the Weyl orbits of the roots
with different length. Thus the root type Lax
pair based on the set of roots itself is
conceptually most suitable for the purpose of
verifying the independence of the coupling
constants.
 For most theories this means two independent coupling
constants, one for the long and the 
other for the short roots potentials.
However, for the model based on the \(BC_n\) root system,
there are three independent coupling constants.

We give two different root type Lax pairs for most of the
untwisted non-simply laced models, one based on the set of long
roots and the other on the set of short roots.
Both give the identical Hamiltonian and  equation of motion.
The list of the Lax pairs is complete in the sense that it
contains all the models with all  four choices of potential
and with and without the spectral parameter, except for the
\(G_2\) model based on the long roots.
In this case new functions satisfying constraints related with
the third order folding are necessary. For
the rational, trigonometric and hyperbolic potentials the
functions given in section 2 satisfy the new constraints, too.
A new set of functions with and without spectral parameter is
obtained for the elliptic potential case.  The actual
verification that these Lax pairs are equivalent with the
canonical equation of motion goes almost parallel with  that of
the root type Lax pairs based on simply laced root systems
\cite{bcs}.
The functions appearing in the root type Lax pairs are the same
for the simply laced and the untwisted non-simply laced cases,
except for the  \(G_2\) case mentioned above.
So we only give the explicit forms of the Lax pairs
for each of the Calogero-Moser models
based on untwisted non-simply laced root systems.

So  far the Lax pairs for untwisted non-simply laced  models were
given in some of the minimal representations only
\cite{OP1,DHPh,bcs}. The situation was a bit confusing: 
the allowed number of independent 
coupling constants can be different
for two different representations of the minimal 
type Lax pair for one and the same
theory.  Now we have the universal
root type Lax pairs  for the untwisted non-simply laced
models with independent coupling constants.

In most cases we normalise the (length)\(^2=2\) for the
long roots, except for  the \(C_n\) and \(BC_n\) system in which
(length)\(^2=4\) is used. They are denoted by subscript $L$. 
For \(G_2\) case we choose to normalise (length)\(^2=3\) for the
long roots and (length)\(^2=1\) for the short roots only for
convenience. The coupling
constant \(g\) without suffix is  reserved for the long roots,
except for the \(C_n\) and
\(BC_n\) systems in which it is used for the short and
middle roots coupling and the long root
coupling constant is denoted by
\(g_L\) specifically. The short root coupling is denoted by
\(g_s\).

\subsection{\(B_n\) model}
The set of \(B_n\) roots consists of two parts, 
long roots and short roots:
\begin{equation}
   \Delta_{B_n}=\Delta\cup\Delta_s,
\end{equation}
in which the roots are conveniently expressed in terms of
an orthonormal basis of \({\mathbb R}^n\):
\begin{eqnarray}
   \Delta&=&\{\alpha,\beta,\gamma,\ldots,\}=\{\pm e_j\pm e_k:
   \quad j,k=1,\ldots,n\},
   \quad 2n(n-1)\ \mbox{roots},
   \nonumber\\
   \Delta_s&=&\{\lambda,\mu,\nu,\ldots, \}=\{\pm e_j: \qquad \quad
   j=1,\ldots,n\},
   \quad 2n\ \mbox{roots.}
   \label{bnroots}
\end{eqnarray}
\noindent From this we know the root difference pattern:
\begin{equation}
   B_n:\qquad \mbox{short root}
   - \mbox{short root}=\left\{
   \begin{array}{l}
      \mbox{long root}\\
      2\times \mbox{short root}\\
      \mbox{non-root}
   \end{array}
   \right.
   \label{bnshrtshrt}
\end{equation}
and
\begin{equation}
   B_n:\qquad \mbox{long root}
   - \mbox{long root}=\left\{
   \begin{array}{l}
      \mbox{long root}\\
      2\times \mbox{long root}\\
      2\times \mbox{short root}\\
      \mbox{non-root}
   \end{array}
   \right.
   \label{bnlonglong}
\end{equation}
\noindent From this knowledge only we can construct the root type Lax
pair for the \(B_n\) model by following the recipe of
the root type Lax pair for simply laced models.

\subsubsection{Root type Lax pair for untwisted \(B_n\) 
model based on short roots
\(\Delta_s\)}
The Lax pair is given in terms of the short roots. %
The matrix elements of \(L_s\) and \(M_s\)  are
labeled  by indices
\(\mu,\nu\) etc.:
\begin{eqnarray}
   	L_s(q,p,\xi) & = & p\cdot H + X + X_{d},\quad \nonumber\\
   	M_s(q,\xi) & = & D+Y+Y_{d}.
   	\label{eq:bnshLaxform}
\end{eqnarray}
Here \(X\) and \(Y\) correspond to the part
of ``short root $-$ short root $=$ long root'' of
(\ref{bnshrtshrt}):
\begin{equation}
   	X=ig\sum_{\alpha\in\Delta}x(\alpha\cdot
    q, \xi)E(\alpha),\quad
   	Y=ig\sum_{\alpha\in\Delta}y(\alpha\cdot
    q, \xi)E(\alpha),\quad
   	E(\alpha)_{\mu \nu}=\delta_{\mu-\nu,\alpha},
   	\label{eq:bnshXYdef}
\end{equation}
and $X_d$ and $Y_d$ correspond to 
``short root $-$ short root $=$ 2\(\times\) short
root'' of (\ref{bnshrtshrt}):
\begin{equation}
   	X_d=2ig_s\sum_{\lambda\in\Delta_s}
    x_{d}(\lambda\cdot q, \xi)E_{d}(\lambda),\quad
   	Y_d=ig_s\sum_{\lambda\in\Delta_s}
    y_{d}(\lambda\cdot q, \xi)E_{d}(\lambda),\quad
   	E_{d}(\lambda)_{\mu \nu}=\delta_{\mu-\nu,2\lambda}.
   	\label{eq:bnshXYrdef}
\end{equation}
The diagonal parts of \(L_s\) and \(M_s\) are given by
\begin{equation}
   	H_{\mu \nu}=\mu \delta_{\mu, \nu},\quad
   	D_{\mu \nu}= \delta_{\mu, \nu}D_{\mu},\quad
   	D_{\mu}=-i\left(g_s\,z(\mu\cdot q)
    +g\sum_{\gamma\in\Delta,\  
   	\gamma\cdot\mu=1}z(\gamma\cdot q)\right).
   	\label{eq:bnshHD}
\end{equation}
The functions \(x,y,z\) and \(x_d,y_d,z_d\) are the same as 
those given in section 2. It is easy to verify
\begin{equation}
   Tr(L_s^2)=4{\cal H}_{B_n},
   \label{bnshl2}
\end{equation}
in which the \(B_n\) Hamiltonian is given by
\begin{equation}
   	{\cal H}_{B_n}={1\over2}p^2-{g^2\over2}
   \sum_{\alpha\in\Delta}
   	x(\alpha\cdot q)x(-\alpha\cdot q)-
   {g_s^2}\sum_{\lambda\in\Delta_s}
   	x(\lambda\cdot q)x(-\lambda\cdot q).
   	\label{eq:bngenham}
\end{equation}

\subsubsection{Root type Lax pair for untwisted \(B_n\) 
model based on long roots
\(\Delta\)}
The Lax pair is given in terms of the long roots. %
The matrix elements of \(L_l\) and \(M_l\)  are
labeled  by indices
\(\alpha,\beta\) etc.:
\begin{eqnarray}
   	L_l(q,p,\xi) & = & p\cdot H + X + X_{d}+X_s,
   \quad \nonumber\\
   	M_l(q,\xi) & = & D+Ds+Y+Y_{d}+Y_s.
   	\label{eq:bnlnLaxform}
\end{eqnarray}
Here \(X\) and \(Y\) correspond to the part
of ``long root $-$ long root $=$ long root'' of
(\ref{bnlonglong}):
\begin{equation}
   	X=ig\sum_{\alpha\in\Delta}x(\alpha\cdot
    q, \xi)E(\alpha),\quad
   	Y=ig\sum_{\alpha\in\Delta}y(\alpha\cdot
    q, \xi)E(\alpha),\quad
   	E(\alpha)_{\beta \gamma}=\delta_{\beta-\gamma,\alpha},
   	\label{eq:bnlnXYdef}
\end{equation}
and $X_d$ and $Y_d$ correspond to 
``long root $-$ long root $=$ 2\(\times\) long root''
of (\ref{bnlonglong}):
\begin{equation}
   	X_d=2ig\sum_{\alpha\in\Delta}
    x_{d}(\alpha\cdot q, \xi)E_{d}(\alpha),\quad
   	Y_d=ig\sum_{\alpha\in\Delta}
    y_{d}(\alpha\cdot q, \xi)E_{d}(\alpha),\quad
   	E_{d}(\alpha)_{\beta \gamma}
    =\delta_{\beta-\gamma,2\alpha}.
   	\label{eq:bnlnXYrdef}
\end{equation}
An additional term in \(L_l\) (\(M_l\)),
$X_s$ ($Y_s$) corresponds to 
``long root $-$ long root $=$ 2\(\times\) short root''
of (\ref{bnlonglong}):
\begin{equation}
   	X_s=2ig_s\sum_{\lambda\in\Delta_s}
    x_{d}(\lambda\cdot q, \xi)E_{d}(\lambda),\quad
   	Y_s=ig_s\sum_{\lambda\in\Delta_s}
    y_{d}(\lambda\cdot q, \xi)E_{d}(\lambda),\quad
   	E_{d}(\lambda)_{\beta \gamma}
    =\delta_{\beta-\gamma,2\lambda}.
   	\label{eq:bnlnXYsdef}
\end{equation}
The diagonal parts of \(L_l\) and \(M_l\) are given by
\begin{equation}
   	H_{\beta \gamma}=\beta \delta_{\beta, \gamma},\quad
   	D_{\beta \gamma}= \delta_{\beta, \gamma}D_{\beta},\quad
   	D_{\beta}=-ig\left(z(\beta\cdot q)+\sum_{\kappa\in\Delta,\  
   	\kappa\cdot\beta=1}z(\kappa\cdot q)\right),
   \label{eq:bnlnHD}
\end{equation}
and
\begin{equation}
    (Ds)_{\beta \gamma}= \delta_{\beta,
    \gamma}(Ds)_{\beta},\quad
   	(Ds)_{\beta}=-ig_s\sum_{\lambda\in\Delta_s,\  
   	\beta\cdot\lambda=1}z(\lambda\cdot q).
   	\label{eq:bnlndsHD}
\end{equation}
The functions \(x,y,z\) 
and \(x_d,y_d,z_d\) are  also the same as 
are given in section 2. 
It is easy to verify that
\begin{equation}
   Tr(L_l^2)=8(n-1){\cal H}_{B_n},
   \label{bnlnl2}
\end{equation}
in which the \(B_n\) Hamiltonian is the same as given above
(\ref{eq:bngenham}). In both cases the reduction from
\(D_{n+1}\) fixes \(g_s=g\). Needless to say, the
consistency of the root type Lax pairs
(\ref{eq:bnshLaxform}) and  (\ref{eq:bnlnLaxform}) 
does not depend on the explicit representation of the
roots in terms of the orthonormal basis
(\ref{bnroots}). This remark applies to the other
models as well.

\subsection{\(C_n\) model}
The set of \(C_n\) roots consists of two parts, 
long roots and short roots:
\begin{equation}
   \Delta_{C_n}=\Delta_L\cup\Delta,
\end{equation}
in which the roots are conveniently expressed in terms of
an orthonormal basis of \({\mathbb R}^n\):
\begin{eqnarray}
   \Delta_L&=&\{\Xi,\Upsilon,\Omega,\ldots,\}=\{\pm 2e_j:
   \qquad \quad j=1,\ldots,n\},
   \quad 2n\ \mbox{roots.},
   \label{cnLnroots}\\
   \Delta&=&\{\alpha,\beta,\gamma,\ldots, \}=
   \{\pm e_j\pm e_k: \quad j,k=1,\ldots,n\},\quad
   2n(n-1)\ \mbox{roots}.
   \label{cnshroots}
\end{eqnarray}
The root difference pattern is
\begin{equation}
   C_n:\qquad \mbox{short root}
   - \mbox{short root}=\left\{
   \begin{array}{l}
      \mbox{short root}\\
      2\times \mbox{short root}\\
      \mbox{long root}\\
      \mbox{non-root}
   \end{array}
   \right.
   \label{cnshrtshrt}
\end{equation}
and
\begin{equation}
   C_n:\qquad \mbox{long root}
   - \mbox{long root}=\left\{
   \begin{array}{l}
      2\times \mbox{long root}\\
      2\times \mbox{short root}\\
      \mbox{non-root}
   \end{array}
   \right.
   \label{cnlonglong}
\end{equation}
\noindent From this knowledge only we can construct the root type Lax
pair for the \(C_n\) model by following the recipe of
the root type Lax pair for simply laced models.

\subsubsection{Root type Lax pair for untwisted \(C_n\) model
based on short roots
\(\Delta\)}
The Lax pair is given in terms of short roots.
The matrix elements of \(L_s\) and \(M_s\)  are
labeled  by indices
\(\beta,\gamma\) etc.:
\begin{eqnarray}
   	L_s(q,p,\xi) & = & p\cdot H + X + X_{d}+X_L,
   \nonumber\\
   	M_s(q,\xi) & = & D+D_L+Y+Y_{d}+X_L,
   	\label{eq:cnshLaxform}
\end{eqnarray}
Here \(X\) and \(Y\) correspond to the part
of ``short root $-$ short root $=$ short root'' of
(\ref{cnshrtshrt}):
\begin{equation}
   	X=ig\sum_{\alpha\in\Delta}x(\alpha\cdot
    q, \xi)E(\alpha),\quad
   	Y=ig\sum_{\alpha\in\Delta}y(\alpha\cdot
    q, \xi)E(\alpha),\quad
   	E(\alpha)_{\beta \gamma}=\delta_{\beta-\gamma,\alpha},
   	\label{eq:cnshXYdef}
\end{equation}
and $X_d$ and $Y_d$ correspond to 
``short root $-$ short root $=$ 2\(\times\) short
root'' of (\ref{cnshrtshrt}):
\begin{equation}
   	X_d=2ig\sum_{\alpha\in\Delta}
    x_{d}(\alpha\cdot q, \xi)E_{d}(\alpha),\quad
   	Y_d=ig\sum_{\alpha\in\Delta}
    y_{d}(\alpha\cdot q, \xi)E_{d}(\alpha),\quad
   	E_{d}(\alpha)_{\beta \gamma}=\delta_{\beta-\gamma,2\alpha}.
   	\label{eq:cnshXYrdef}
\end{equation}
An additional term in \(L_s\) (\(M_s\)),
$X_L$ ($Y_L$) corresponds to 
``short root $-$ short root $=$ long root''
of (\ref{cnlonglong}):
\begin{equation}
   	X_L=ig_L\sum_{\Xi\in\Delta_L}
    x(\Xi\cdot q, \xi)E(\Xi),\quad
   	Y_L=ig_L\sum_{\Xi\in\Delta_L}
    y(\Xi\cdot q, \xi)E(\Xi),\quad
   	E(\Xi)_{\beta
\gamma}=\delta_{\beta-\gamma,\Xi}.
   	\label{eq:cnlnXYLdef}
\end{equation}
The diagonal parts of \(L_s\) and \(M_s\) are given by
\begin{equation}
   	H_{\beta \gamma}=\beta \delta_{\beta, \gamma},\quad
   	D_{\beta \gamma}= \delta_{\beta, \gamma}D_{\beta},\quad
   	D_{\beta}=-ig\left(z(\beta\cdot q)
   +\sum_{\kappa\in\Delta,\  
   	\kappa\cdot\beta=1}z(\kappa\cdot q)\right),
   \label{eq:cnshHD}
\end{equation}
and
\begin{equation}
   (D_L)_{\beta \gamma}= \delta_{\beta,
   \gamma}(D_L)_{\beta},\quad
   	(D_L)_{\beta}=-ig_L\sum_{\Upsilon\in\Delta_L,\  
   	\beta\cdot\Upsilon=2}z(\Upsilon\cdot q).
   	\label{eq:cnshdLHD}
\end{equation}
The functions \(x,y,z\) and \(x_d,y_d,z_d\) are the same as 
are given in section 2. It is easy to verify
\begin{equation}
   Tr(L_s^2)=8(n-1){\cal H}_{C_n},
   \label{cnshl2}
\end{equation}
in which \(C_n\) Hamiltonian is given by
\begin{equation}
   	{\cal H}_{C_n}={1\over2}p^2-{g^2\over2}
    \sum_{\alpha\in\Delta}
   	x(\alpha\cdot q)x(-\alpha\cdot
    q)-{g_L^2\over4}\sum_{\Xi\in\Delta_L}
   	x(\Xi\cdot q)x(-\Xi\cdot q).
   	\label{eq:cngenham}
\end{equation}

\subsubsection{Root type Lax pair for untwisted
\(C_n\) model based on long roots
\(\Delta_L\)}
The Lax pair is given in terms of long roots.
The matrix elements of \(L_L\) and \(M_L\)  are
labeled  by indices
 \(\Upsilon,\Omega\)
etc.:
\begin{eqnarray}
   	L_L(q,p,\xi) & = & p\cdot H +  X_{d}+X_s, \nonumber\\
   	M_L(q,\xi) & = & D+Y_{d}+Y_s.
   	\label{eq:cnlnLaxform}
\end{eqnarray}
Here \(X_d\) and \(Y_d\) correspond to the part
of ``long root $-$ long root $=2\times$ long root'' of
(\ref{cnlonglong}):
\begin{equation}
   	X_d=2ig_L\sum_{\Xi\in\Delta_L}x_d(\Xi\cdot
   q, \xi)E_d(\Xi),\quad
   	Y_d=ig_L\sum_{\Xi\in\Delta_L}y_d(\Xi\cdot
   q, \xi)E_d(\Xi),\quad
   	E_d(\Xi)_{\Upsilon \Omega}=\delta_{\Upsilon-\Omega,2\Xi},
   	\label{eq:cnlnXYdef}
\end{equation}
and $X_s$ and $Y_s$ correspond to 
``long root $-$ long root $=$ 2\(\times\) short root''
of (\ref{cnlonglong}):
\begin{equation}
   	X_s=2ig\sum_{\alpha\in\Delta}
   x_{d}(\alpha\cdot q, \xi)E_{d}(\alpha),\quad
   	Y_s=ig\sum_{\alpha\in\Delta}
   y_{d}(\alpha\cdot q, \xi)E_{d}(\alpha),\quad
   	E_{d}(\alpha)_{\Upsilon
   \Omega}=\delta_{\Upsilon-\Omega,2\alpha}.
   	\label{eq:cnlnXYrdef}
\end{equation}
The diagonal parts of \(L_L\) and \(M_L\) are given by
\begin{equation}
   	H_{\Upsilon \Omega}=\Upsilon \delta_{\Upsilon, \Omega},\quad
   	D_{\Upsilon \Omega}= \delta_{\Upsilon,
   \Omega}D_{\Upsilon},\quad
   	D_{\Upsilon}=-i\left(g_L\,z(\Upsilon\cdot
   q)+g\sum_{\kappa\in\Delta,\  
   	\kappa\cdot\Upsilon=2}z(\kappa\cdot q)\right).
   \label{eq:cnlnHD}
\end{equation}
Since only the functions \(x_d,y_d\) appear and no
functions \(x,y\) appear in the Lax pair, we can safely use
\(x,y,z\),  which are used in the minimal type Lax pairs,
in place of \(x_d,y_d,z_d\). 
It should be noted that the set of long roots
(\ref{cnLnroots}) is 2 times the set of vector weights of
\(C_n\):
\[
   \Lambda=\{\pm e_j:\ j=1,\ldots,n\}.
\]
In fact the above \(L_L\) matrix is twice 
 \(L\) for the vector representation with two independent
coupling constants \cite{OP1,bcs} (with proper
identification):
\[
   L_L=2L_{vec},\quad M_L=M_{vec}.
\]
In other words the root type Lax pair based on \(C_n\)
Long
roots is equivalent with the vector representation Lax pair.
This explains why two independent coupling constants are
allowed in the vector representation Lax pair of the \(C_n\) 
model \cite{OP1}.

\subsection{\(F_4\) model}
The set of \(F_4\) roots consists of two parts, 
long  and short roots:
\begin{equation}
   \Delta_{F_4}=\Delta\cup\Delta_s,
\end{equation}
in which the roots are conveniently expressed in terms of
an orthonormal basis of \({\mathbb R}^4\):
\begin{eqnarray}
   \Delta&=&\{\alpha,\beta,\gamma,\ldots,\}=\{\pm e_j\pm e_k:
   \quad j,k=1,\ldots,4\},
   \quad 24\ \mbox{roots},
   \nonumber\\
   \Delta_s&=&\{\lambda,\mu,\nu,\ldots, \}=\{\pm e_j, 
   {1\over2}(\pm e_1\pm e_2\pm e_3\pm e_4):
   \ j=1,\ldots,4\},
   \quad 24\ \mbox{roots.}
   \label{f4roots}
\end{eqnarray}
The set of long roots has the same structure as the \(D_4\)
roots and the set of short roots has the same structure as
the union of \(D_4\) vector, spinor and anti-spinor weights.
\noindent From this we know the root difference pattern:
\begin{equation}
   F_4:\qquad \mbox{short root}
   - \mbox{short root}=\left\{
   \begin{array}{l}
      \mbox{long root}\\
       \mbox{short root}\\
      2\times \mbox{short root}\\
      \mbox{non-root}
   \end{array}
   \right.
   \label{f4shrtshrt}
\end{equation}
and
\begin{equation}
   F_4:\qquad \mbox{long root}
   - \mbox{long root}=\left\{
   \begin{array}{l}
      \mbox{long root}\\
      2\times \mbox{long root}\\
      2\times \mbox{short root}\\
      \mbox{non-root}
   \end{array}
   \right.
   \label{f4longlong}
\end{equation}
\noindent From this knowledge only we can construct the root type Lax
pair for \(F_4\) model by following the same recipe as above.

\subsubsection{Root type Lax pair for untwisted \(F_4\) 
model based on short roots
\(\Delta_s\)}
The Lax pair is given in terms of short roots. 
The matrix elements of \(L_s\) and \(M_s\)  are
labeled  by indices
\(\mu,\nu\) etc.: 
\begin{eqnarray}
   	L_s(q,p,\xi) & = & p\cdot H + X + X_{d}+X_l, \nonumber\\
   	M_s(q,\xi) & = & D+D_l+Y+Y_{d}+Y_l,
   	\label{eq:f4shLaxform}
\end{eqnarray}
Here \(X\) and \(Y\) correspond to the part
of ``short root $-$ short root $=$ short root'' of
(\ref{f4shrtshrt}):
\begin{equation}
   	X=ig_s\sum_{\lambda\in\Delta_s}x(\lambda\cdot
    q, \xi)E(\lambda),\quad
   	Y=ig_s\sum_{\lambda\in\Delta_s}y(\lambda\cdot
    q, \xi)E(\lambda),\quad
   	E(\lambda)_{\mu \nu}=\delta_{\mu-\nu,\lambda},
   	\label{eq:f4shXYdef}
\end{equation}
and $X_d$ and $Y_d$ correspond to 
``short root $-$ short root $=$ 2\(\times\) short
root'' of (\ref{f4shrtshrt}):
\begin{equation}
   	X_d=2ig_s\sum_{\lambda\in\Delta_s}
    x_{d}(\lambda\cdot q, \xi)E_{d}(\lambda),\quad
   	Y_d=ig_s\sum_{\lambda\in\Delta_s}
    y_{d}(\lambda\cdot q, \xi)E_{d}(\lambda),\quad
   	E_{d}(\lambda)_{\mu \nu}=\delta_{\mu-\nu,2\lambda}.
   	\label{eq:f4shXYrdef}
\end{equation}
The additional terms \(X_l\) and \(Y_l\) correspond to 
``short root $-$ short root $=$  long
root'' of (\ref{f4shrtshrt}):
\begin{equation}
   	X_l=ig\sum_{\alpha\in\Delta}x(\alpha\cdot
    q, \xi)E(\alpha),\quad
   	Y_l=ig\sum_{\alpha\in\Delta}y(\alpha\cdot
    q, \xi)E(\alpha),\quad
   	E(\alpha)_{\mu \nu}=\delta_{\mu-\nu,\alpha}.
   	\label{eq:f4shXYlndef}
\end{equation}
The diagonal parts of \(L_s\) and \(M_s\) are given by
\begin{equation}
   	H_{\mu \nu}=\mu \delta_{\mu, \nu},\quad
   	D_{\mu \nu}= \delta_{\mu, \nu}D_{\mu},\quad
   	D_{\mu}=-ig_s\left(\,z(\mu\cdot
   q)+\sum_{\lambda\in\Delta_s,\  
   	\lambda\cdot\mu=1/2}z(\lambda\cdot q)\right),
   	\label{eq:f4shHD}
\end{equation}
and
\begin{equation}
   (D_l)_{\mu \nu}= \delta_{\mu,
   \nu}(D_l)_{\mu},\quad
   	(D_l)_{\mu}=-ig\sum_{\alpha\in\Delta,\  
   	\alpha\cdot\mu=1}z(\alpha\cdot q).
   	\label{eq:f4shdLHD}
\end{equation}
The functions \(x,y,z\) and \(x_d,y_d,z_d\) are the same as 
are given in section 2. It is easy to verify that
\begin{equation}
   Tr(L_s^2)=12{\cal H}_{F_4},
   \label{f4shl2}
\end{equation}
in which the \(F_4\) Hamiltonian is given by
\begin{equation}
   	{\cal H}_{F_4}={1\over2}p^2-{g^2\over2}
   \sum_{\alpha\in\Delta}
   	x(\alpha\cdot q)x(-\alpha\cdot q)-
   {g_s^2}\sum_{\lambda\in\Delta_s}
   	x(\lambda\cdot q)x(-\lambda\cdot q).
   	\label{eq:f4genham}
\end{equation}
It should be noted that  this has the same general structure
as the Hamiltonian of the \(B_n\) theory (\ref{eq:bngenham}).

\subsubsection{Root type Lax pair for untwisted \(F_4\) 
model based on long roots
\(\Delta\)}
The Lax pair is given in terms of long roots. 
The general structure of this Lax pair is essentially the same as
that of the \(B_n\) theory, since the pattern of the long root--
long root  (\ref{f4longlong}) is the same as that
of \(B_n\) (\ref{bnlonglong}). This reflects the universal
nature of the root type Lax pairs.
So we list the general form only without further explanation.
 They are
matrices with indices
\(\beta,\gamma\) etc.:
\begin{eqnarray}
   	L_l(q,p,\xi) & = & p\cdot H + X + X_{d}+X_s, \nonumber\\
   	M_l(q,\xi) & = & D+Ds+Y+Y_{d}+Y_s.
   	\label{eq:f4lnLaxform}
\end{eqnarray}
\begin{equation}
   	X=ig\sum_{\alpha\in\Delta}x(\alpha\cdot
    q, \xi)E(\alpha),\quad
   	Y=ig\sum_{\alpha\in\Delta}y(\alpha\cdot
    q, \xi)E(\alpha),\quad
   	E(\alpha)_{\beta \gamma}=\delta_{\beta-\gamma,\alpha}.
   	\label{eq:f4lnXYdef}
\end{equation}
\begin{equation}
   	X_d=2ig\sum_{\alpha\in\Delta}
    x_{d}(\alpha\cdot q, \xi)E_{d}(\alpha),\quad
   	Y_d=ig\sum_{\alpha\in\Delta}
    y_{d}(\alpha\cdot q, \xi)E_{d}(\alpha),\quad
   	E_{d}(\alpha)_{\beta \gamma}
    =\delta_{\beta-\gamma,2\alpha}.
   	\label{eq:f4lnXYrdef}
\end{equation}
\begin{equation}
  	X_s=2ig_s\sum_{\lambda\in\Delta_s}
   x_{d}(\lambda\cdot q, \xi)E_{d}(\lambda),\quad
  	Y_s=ig_s\sum_{\lambda\in\Delta_s}
   y_{d}(\lambda\cdot q, \xi)E_{d}(\lambda),\quad
  	E_{d}(\lambda)_{\beta \gamma}
   =\delta_{\beta-\gamma,2\lambda}.
  	\label{eq:f4lnXYsdef}
\end{equation}
The diagonal parts of \(L_l\) and \(M_l\) are given by
\begin{equation}
   	H_{\beta \gamma}=\beta \delta_{\beta, \gamma},\quad
   	D_{\beta \gamma}= \delta_{\beta, \gamma}D_{\beta},\quad
   	D_{\beta}=-ig\left(z(\beta\cdot q)
    +\sum_{\kappa\in\Delta,\  
   	\kappa\cdot\beta=1}z(\kappa\cdot q)\right),
   \label{eq:f4lnHD}
\end{equation}
and
\begin{equation}
    (Ds)_{\beta \gamma}= \delta_{\beta,
\gamma}(Ds)_{\beta},\quad
   	(Ds)_{\beta}=-ig_s\sum_{\lambda\in\Delta_s,\  
   	\beta\cdot\lambda=1}z(\lambda\cdot q).
   	\label{eq:f4lndsHD}
\end{equation}
The functions \(x,y,z\) and \(x_d,y_d,z_d\) are  
also the same as 
are given in section 2. 
It is easy to verify that
\begin{equation}
   Tr(L_l^2)=24{\cal H}_{F_4},
   \label{f4lnl2}
\end{equation}
in which the \(F_4\) Hamiltonian is the same as given above
(\ref{eq:bngenham}). In both cases the reduction from
\(E_{6}\) fixes \(g_s=g\).

\subsection{\(G_2\) model}
The set of \(G_2\) roots consists of two parts, 
long  and short roots:
\begin{equation}
   \Delta_{G_2}=\Delta\cup\Delta_s,
\end{equation}
in which the roots are conveniently expressed in terms of
an orthonormal basis of \({\mathbb R}^2\):
\begin{eqnarray}
   \Delta&=&\{\alpha,\beta,\gamma,\ldots,\}=
   \{\pm(-3e_1+\sqrt3e_2)/2,
   \pm(3e_1+\sqrt3e_2)/2,
   \pm\sqrt3 e_2\},
   \quad 6\ \mbox{roots},
   \nonumber\\
   \Delta_s&=&\{\lambda,\mu,\nu,\ldots, \}=\{\pm e_1, 
   \pm(-e_1+\sqrt3e_2)/2,\pm(e_1+\sqrt3e_2)/2\},
   \quad 6\ \mbox{roots.}
   \label{g2roots}
\end{eqnarray}
The sets of long and short roots have the same structure as the
\(A_2\) roots, scaled [(long root)\(^2:\ \)(short
root)\(^2=3:1\)] and rotated
\({\pi/6}\). The root difference pattern is:
\begin{equation}
   G_2:\qquad \mbox{short root}
   - \mbox{short root}=\left\{
   \begin{array}{l}
      \mbox{long root}\\
       \mbox{short root}\\
      2\times \mbox{short root}\\
      \mbox{non-root}
   \end{array}
   \right.
   \label{g2shrtshrt}
\end{equation}
\begin{equation}
   G_2:\qquad \mbox{long root}
   - \mbox{long root}=\left\{
   \begin{array}{l}
      \mbox{long root}\\
      2\times \mbox{long root}\\
      3\times \mbox{short root}\\
      \mbox{non-root}
   \end{array}
   \right.
   \label{g2longlong}
\end{equation}
The appearance of \(3\times\) short root in (\ref{g2longlong})
is a new feature. 

\subsubsection{Root type Lax pair for untwisted \(G_2\) 
model based on short roots
\(\Delta_s\)}
The Lax pair is given in terms of short roots. 
The general structure of this Lax pair is essentially the same as
that of \(F_4\) theory, since the pattern of the short root--
short root  (\ref{g2shrtshrt}) is the same as that of
the \(F_4\) (\ref{f4shrtshrt}). So we list the general form
only without further explanation.
The matrix elements of \(L_s\) and \(M_s\)  are
labeled  by indices
\(\mu,\nu\) etc.:
\begin{eqnarray}
   	L_s(q,p,\xi) & = & p\cdot H + X + X_{d}+X_l, \nonumber\\
   	M_s(q,\xi) & = & D+D_l+Y+Y_{d}+Y_l,
   	\label{eq:g2shLaxform}
\end{eqnarray}
\begin{equation}
   	X=ig_s\sum_{\lambda\in\Delta_s}x(\lambda\cdot
   q, \xi)E(\lambda),\quad
   	Y=ig_s\sum_{\lambda\in\Delta_s}y(\lambda\cdot
   q, \xi)E(\lambda),\quad
   	E(\lambda)_{\mu \nu}=\delta_{\mu-\nu,\lambda}.
   	\label{eq:g2shXYdef}
\end{equation}
\begin{equation}
   	X_d=2ig_s\sum_{\lambda\in\Delta_s}
   x_{d}(\lambda\cdot q, \xi)E_{d}(\lambda),\quad
   	Y_d=ig_s\sum_{\lambda\in\Delta_s}
   y_{d}(\lambda\cdot q, \xi)E_{d}(\lambda),\quad
   	E_{d}(\lambda)_{\mu \nu}=\delta_{\mu-\nu,2\lambda}.
   	\label{eq:g2shXYrdef}
\end{equation}
\begin{equation}
   	X_l=ig\sum_{\alpha\in\Delta}x(\alpha\cdot
   q, \xi)E(\alpha),\quad
   	Y_l=ig\sum_{\alpha\in\Delta}y(\alpha\cdot
   q, \xi)E(\alpha),\quad
   	E(\alpha)_{\mu \nu}=\delta_{\mu-\nu,\alpha}.
   	\label{eq:g2shXYlndef}
\end{equation}
The diagonal parts of \(L_s\) and \(M_s\) are given by
\begin{equation}
   	H_{\mu \nu}=\mu \delta_{\mu, \nu},\quad
   	D_{\mu \nu}= \delta_{\mu, \nu}D_{\mu},\quad
   	D_{\mu}=-ig_s\left(\,z(\mu\cdot
    q)+\sum_{\lambda\in\Delta_s,\  
   	\lambda\cdot\mu=1/2}z(\lambda\cdot q)\right),
   	\label{eq:g2shHD}
\end{equation}
and
\begin{equation}
   (D_l)_{\mu \nu}= \delta_{\mu,
   \nu}(D_l)_{\mu},\quad
   	(D_l)_{\mu}=-ig\sum_{\alpha\in\Delta,\  
   	\alpha\cdot\mu=3/2}z(\alpha\cdot q).
   	\label{eq:g2shdLHD}
\end{equation}
The functions \(x,y,z\) and \(x_d,y_d,z_d\) are the same as 
are given in section 2. It is easy to verify that
\begin{equation}
   Tr(L_s^2)=6{\cal H}_{G_2},
   \label{g2shl2}
\end{equation}
in which the \(G_2\) Hamiltonian is given by
\begin{equation}
   	{\cal H}_{G_2}={1\over2}p^2-{g^2\over3}
   \sum_{\alpha\in\Delta}
   	x(\alpha\cdot q)x(-\alpha\cdot q)-
   {g_s^2}\sum_{\lambda\in\Delta_s}
   	x(\lambda\cdot q)x(-\lambda\cdot q).
   	\label{eq:g2genham}
\end{equation}

\subsubsection{Root type Lax pair for untwisted \(G_2\) 
model based on long roots
\(\Delta\)}
This Lax pair is different from the others because of the
`triple root' term in (\ref{g2longlong}).
The matrix elements of \(L_l\) and \(M_l\)  are
labeled  by indices
\(\beta,\gamma\) etc.:
\begin{eqnarray}
   	L_l(q,p,\xi) & = & p\cdot H + X + X_{d}+X_t, \nonumber\\
   	M_l(q,\xi) & = & D+Dt+Y+Y_{d}+Y_t.
   	\label{eq:g2lnLaxform}
\end{eqnarray}
The terms \(X,Y\) and \(X_d,Y_d\) are the same as before:
\begin{equation}
   	X=ig\sum_{\alpha\in\Delta}x(\alpha\cdot
   q, \xi)E(\alpha),\quad
   	Y=ig\sum_{\alpha\in\Delta}y(\alpha\cdot
   q, \xi)E(\alpha),\quad
   	E(\alpha)_{\beta \gamma}=\delta_{\beta-\gamma,\alpha}.
   	\label{eq:g2lnXYdef}
\end{equation}
\begin{equation}
   	X_d=2ig\sum_{\alpha\in\Delta}
   x_{d}(\alpha\cdot q, \xi)E_{d}(\alpha),\quad
   	Y_d=ig\sum_{\alpha\in\Delta}
   y_{d}(\alpha\cdot q, \xi)E_{d}(\alpha),\quad
   	E_{d}(\alpha)_{\beta \gamma}
   =\delta_{\beta-\gamma,2\alpha}.
   	\label{eq:g2lnXYrdef}
\end{equation}
The terms \(X_t\) and \(Y_t\) are associated with the 
`triple root' with new functions \(x_t\) and \(y_t\):
\begin{equation}
   	X_t=3ig_s\sum_{\lambda\in\Delta_s}
   x_{t}(\lambda\cdot q, \xi)E_{t}(\lambda),\quad
   	Y_t=ig_s\sum_{\lambda\in\Delta_s}
   y_{t}(\lambda\cdot q, \xi)E_{t}(\lambda),\quad
   	E_{t}(\lambda)_{\beta \gamma}
   =\delta_{\beta-\gamma,3\lambda}.
   	\label{eq:g2lnXYtdef}
\end{equation}
The diagonal parts of \(L_l\) and \(M_l\) are given by
\begin{equation}
   	H_{\beta \gamma}=\beta \delta_{\beta, \gamma},\quad
   	D_{\beta \gamma}= \delta_{\beta, \gamma}D_{\beta},\quad
   	D_{\beta}=-ig\left(z(\beta\cdot q)
   +\sum_{\kappa\in\Delta,\  
   	\kappa\cdot\beta=3/2}z(\kappa\cdot q)\right),
   \label{eq:g2lnHD}
\end{equation}
and
\begin{equation}
    (Dt)_{\beta \gamma}= \delta_{\beta,
\gamma}(Dt)_{\beta},\quad
   	(Dt)_{\beta}=-ig_s\sum_{\lambda\in\Delta_s,\  
   	\beta\cdot\lambda=3/2}z(\lambda\cdot q).
   	\label{eq:g2lndtHD}
\end{equation}
The pairs of functions \(\{x,y\}\), 
\(\{x_d,y_d\}\) and \(\{x_t,y_t\}\)
should each satisfy the {\em sum rule} (\ref{eq:ident1}).
As in the other cases  \(\{x,y\}\) and \(\{x_d,y_d\}\) should
satisfy the {\em second sum rule}
 (\ref{eq:xydiden}).
There is also a {\em third sum rule} to 
be satisfied by all of these functions:
\begin{eqnarray}
   0&=&x(2u-v)y(u-2v)-x(u-2v)y(2u-v)
   -x(3v)\,y_t(u-2v)+y_t(2u-v)\,x(-3u)\nonumber\\
   &&-2x_d(3u)\,y_t(-u-v)+2y_t(u+v)\,x_d(-3v)
   -3x_t(2u-v)\,y(-3u)+3y(3v)\,x_t(u-2v)\nonumber\\
   &&-3x_t(u+v)\,y_d(-3v)+3y_d(3u)\,x_t(-u-v).
   \label{tripiden}
\end{eqnarray}
For the rational, trigonometric and hyperbolic potentials,
all three sum rules are satisfied by the same set of functions
as before:
\begin{eqnarray}
   	x(t)&=&x_{d}(t)=x_{t}(t)={1\over t},\qquad \quad \quad \
   y(t)=y_{d}(t)=y_{t}(t)=-{1\over{t^2}},
   	\quad z(t)=-{1\over{t^2}},\nonumber\\
   			x(t)&=&x_{d}(t)=x_{t}(t)=a\cot at,
   \qquad y(t)=y_{d}(t)=y_{t}(t)=
   -{a^{2}\over{\sin^2 at}},\nonumber\\
   			&&\hspace{2cm} z(t)=z_{d}(t)=z_{t}(t)
   =-{a^{2}\over{\sin^2 at}},\qquad a: const.
   	\nonumber\\
   		x(t)&=&x_{d}(t)=x_{t}(t)=a\coth at,
   \qquad y(t)=y_{d}(t)=y_{t}(t)=
   -{a^{2}\over{\sinh^2 at}},\nonumber\\
   			&&\hspace{2cm} z(t)=z_{d}(t)=z_{t}(t)
   =-{a^{2}\over{\sinh^2 at}},
   	\label{eq:trifunctionshyp}
\end{eqnarray}	
and the Lax pair (\ref{eq:g2lnLaxform}) is equivalent 
with the canonical
equation of motion.
For the elliptic potential with  spectral parameter, a simple set of
solutions is obtained in analogy with the solutions
(\ref{simplesol}):
\begin{eqnarray}
   	x(t,\xi)&=&{\sigma({\xi/3}-
t)\over{\sigma({\xi/3})\sigma(
   t)}},
   \qquad  y(t,\xi)=
    x(t,\xi)\left[\zeta(t-\xi/3)-\zeta(t)\right],
   \nonumber\\ 
   &&\hspace{3.0cm} z(t,\xi)=-\left[\wp(
t)-\wp({\xi/3})\right],\nonumber\\
   x_{d}(t,\xi) &=& {\sigma(2\xi/3-
t)\over{\sigma(2\xi/3)\sigma(
   t)}},
   \quad \ y_d(t,\xi)=
    x_d(t,\xi)\left[\zeta(t-2\xi/3)-\zeta(t)\right],
   \nonumber\\  
   &&\hspace{3.0cm}
z_d(t,\xi)=-\left[\wp(t)-\wp(2\xi/3)\right],\nonumber\\
   x_{t}(t,\xi) &=& {\sigma(\xi-t)\over{\sigma(\xi)\sigma(
t)}},
   \qquad \quad y_t(t,\xi)=
    x_t(t,\xi)\left[\zeta(t-\xi)-\zeta(t)\right],
   \nonumber\\  
   &&\hspace{3.0cm} z_t(t,\xi)=-\left[\wp(t)-\wp(\xi)\right].
   \label{simplesolthree}
\end{eqnarray}
The spectral parameter independent functions are obtained by
setting \(\xi=\omega_j\), (\(j=1,2,3\)) with appropriate
exponential factors:
\begin{eqnarray}
	x(t)&=&{\sigma({\omega_j/3}-
t)\over{\sigma({\omega_j/3})\sigma(
   t)}}\,e^{\eta_jt/3},
   \qquad  \quad y(t)=
    x(t)\left[\zeta(t-\omega_j/3)-\zeta(t)+\eta_j/3\right],
   \nonumber\\ 
   &&\hspace{4.5cm}
z(t)=-\left[\wp(t)-\wp({\omega_j/3})\right],
   \nonumber\\
   x_{d}(t) &=&
{\sigma(2\omega_j/3-t)\over{\sigma(2\omega_j/3)
\sigma(t)}}\,e^{2\eta_jt/3},
   \quad \ \, y_d(t)=  
x_d(t)\left[\zeta(t-2\omega_j/3)-\zeta(t)+2\eta_j/3\right],
   \nonumber\\  
   &&\hspace{4.4cm}
z_d(t)=-\left[\wp(t)-\wp(2\omega_j/3)\right],\nonumber\\
   x_{t}(t) &=& {\sigma(\omega_j-
t)\over{\sigma(\omega_j)\sigma(t)}}\,e^{\eta_jt},
   \qquad \quad \quad \, y_t(t)=
    x_t(t)\left[\zeta(t-\omega_j)-\zeta(t)+\eta_j\right],
   \nonumber\\  
   &&\hspace{4.4cm}
z_t(t)=-\left[\wp(t)-\wp(\omega_j)\right].
   \label{simplesolthreeind}
\end{eqnarray}
They are doubly periodic
meromorphic functions and may be
viewed as generalisations of co-\(\wp\) functions.
For example, for \(j=1\),
\(x(t), x_d(t)\) and \(x_t(t)\) have fundamental
periods
\(\{2\omega_1,12\omega_3\}\),
\(\{2\omega_1,6\omega_3\}\) and \(\{2\omega_1,4\omega_3\}\),
respectively.
The properties of these functions  will be discussed in a future
publication. 

\subsection{\(BC_n\) root system Lax 
pair with three independent couplings}
The \(BC_n\) root system consists of three parts, 
long,  middle and short roots:
\begin{equation}
   \Delta_{BC_n}=\Delta_L\cup\Delta\cup\Delta_s,
\end{equation}
in which the roots are conveniently expressed in terms of
an orthonormal basis of \({\mathbb R}^n\):
\begin{eqnarray}
   \Delta_L&=&\{\Xi,\Upsilon,\Omega,\ldots,\}=\{\pm 2e_j:
   \qquad \quad j=1,\ldots,n\},
   \quad 2n\ \mbox{roots},
   \label{bcnLnroots}\\
   \Delta&=&\{\alpha,\beta,\gamma,\ldots, \}=
   \{\pm e_j\pm e_k: \quad j,k=1,\ldots,n\},\quad
   2n(n-1)\ \mbox{roots},
   \label{bcnmdroots}\\
   \Delta_s&=&\{\lambda,\mu,\nu,\ldots, \}
   =\{\pm e_j: \qquad \quad
   j=1,\ldots,n\},
   \quad \quad \ 2n\ \mbox{roots.}
   \label{bcnroots}
\end{eqnarray}
Here we consider the Lax pair based on the middle roots only.
The pattern of middle root-- middle root  is
\begin{equation}
   BC_n:\qquad \mbox{middle root}
   - \mbox{middle root}=\left\{
   \begin{array}{l}
      \mbox{long root}\\
      \mbox{middle root}\\
      2\times \mbox{middle root}\\
      2\times \mbox{short root}\\
      \mbox{non-root}
   \end{array}
   \right.
   \label{bcnmdmd}
\end{equation}
\noindent From this knowledge only we can construct the root type Lax
pair for \(BC_n\) root system: 
\begin{eqnarray}
   	L_m(q,p,\xi) & = & p\cdot H + X + X_{d}+X_L+X_s, \nonumber\\
   	M_m(q,\xi) & = & D+D_L+Y+Y_{d}+Y_L+Ds+Y_s.
   	\label{eq:bcnmdLaxform}
\end{eqnarray}
The matrix elements of \(L_m\) and \(M_m\)  are
labeled  by indices
\(\beta,\gamma\) etc.
Here \(p\cdot H + X + X_{d}+X_L\) (\(D+D_L+Y+Y_{d}+Y_L\)) 
is exactly the same as \(L_s\) (\(M_s\)) matrix
of \(C_n\) models with two coupling constants based on short
roots.
So we give only the terms related with short roots:
An additional term in \(L_m\) (\(M_m\)),
$X_s$ ($Y_s$) corresponds to 
``middle root $-$ middle root $=$ 2\(\times\) short root''
of (\ref{bcnmdmd}):
\begin{equation}
   	X_s=2ig_s\sum_{\lambda\in\Delta_s}
    x_{d}(\lambda\cdot q, \xi)E_{d}(\lambda),\quad
   	Y_s=ig_s\sum_{\lambda\in\Delta_s}
    y_{d}(\lambda\cdot q, \xi)E_{d}(\lambda),\quad
   	E_{d}(\lambda)_{\beta \gamma}=\delta_{\beta-\gamma,2\lambda}.
   	\label{eq:bcnlnXYsdef}
\end{equation}
\begin{equation}
    Ds_{\beta \gamma}= \delta_{\beta, \gamma}Ds_{\beta},\quad
   	Ds_{\beta}=-ig_s\sum_{\lambda\in\Delta_s,\  
   	\beta\cdot\lambda=1}z(\lambda\cdot q).
   	\label{eq:bcnlndsHD}
\end{equation}
The functions \(x,y,z\) and \(x_d,y_d,z_d\) are the same as 
are given in section 2. It is easy to verify that
\begin{equation}
   Tr(L_m^2)=8(n-1){\cal H}_{BC_n},
   \label{bcnshl2}
\end{equation}
in which the \(BC_n\) Hamiltonian  is the \(C_n\) Hamiltonian 
(\ref{eq:cngenham}) plus
the contribution from the short root potential
with ``renormalisation" of the short root coupling
constant:
\begin{eqnarray}
   	{\cal H}_{BC_n}&=&{1\over2}p^2-{g^2\over2}
    \sum_{\alpha\in\Delta}
   	x(\alpha\cdot q)x(-\alpha\cdot
    q)-{g_L^2\over4}\sum_{\Xi\in\Delta_L}
   	x(\Xi\cdot q)x(-\Xi\cdot
q)\nonumber\\
&&\hspace*{3cm}-{\tilde{g}_s}^2\sum_{\lambda\in\Delta_s}
   	x(\lambda\cdot q)x(-\lambda\cdot q),\nonumber\\
&&\hspace*{3cm}{\tilde{g}_s}^2=g_s(g_s+g_L/2).
   	\label{eq:bcngenham}
\end{eqnarray}

\section{Summary and Comments}
Universal Lax pairs for Calogero-Moser models based on
simply laced root systems are presented for all of the four
choices of potentials: the rational,
trigonometric,  hyperbolic and
elliptic, with and without spectral parameter
(section two). These are the root type Lax pairs
and the minimal type Lax pairs. The Calogero-Moser
models based on simply laced root systems have
 discrete symmetries generated by the
automorphisms of the Dynkin diagrams and the extended
Dynkin diagrams of the root system.  By combining the
discrete symmetry arising from the automorphism
of the extended Dynkin diagram with the
periodicity of the elliptic potential, 
Calogero-Moser models for various  twisted
non-simply laced root systems are derived from
those based on simply laced root systems (section
three). The model associated with the affine
Dynkin diagram
\(A_{2n}^{(2)}\)  can be interpreted as
a twisted version of the \(BC_n\) Calogero-Moser model.

The idea of the universal root type Lax pairs is
successfully generalised to all of the untwisted
non-simply laced Calogero-Moser models (section four).
For non-simply laced root systems, there are two kinds of
root type Lax pairs:
one based on the set of long roots, the other on the set of
short roots.
They both contain as many independent coupling constants
as  independent Weyl orbits in the
set of roots.
For the \(BC_n\) root system, this means that there are three
independent coupling constants.
Consistency of the \(G_2\) root type Lax pair based on long roots
requires a new set of functions when the potential
is elliptic. A simple set of these functions is given.

We have not discussed the unified Lax pairs, root as well as
minimal type,
of the twisted non-simply laced Calogero-Moser models
(with independent coupling constants) derived in section three.
This is  an interesting subject because of its connection with
(affine) Toda (lattice or field) theories.

\begin{center}
{\bf ACKNOWLEDGMENTS}
\end{center}
This work is partially supported  by the Grant-in-aid
from the Ministry of Education, Science and Culture,
Priority Area 
``Supersymmetry and unified theory of elementary particles"
(\#707). A.\,J.\,B. is supported by the Japan Society
for the Promotion of Science  and the National Science
Foundation under grant no. 9703595.

\end{document}